\documentclass[12pt]{article}
\usepackage[a4paper, total={7in, 9in}, textwidth=400pt]{geometry}

\usepackage{amsmath}
\usepackage{booktabs} 
\usepackage{caption} 
\usepackage{subcaption} 
\usepackage{graphicx}
\usepackage{pgfplots}
\usepackage[all]{nowidow}
\usepackage[utf8]{inputenc}
\usepackage{tikz}
\usetikzlibrary{er,positioning,bayesnet}
\usepackage{multicol}
\usepackage{algpseudocode,algorithm,algorithmicx}
\usepackage{hyperref}
\usepackage{cleveref}
\usepackage{float}
\usepackage{placeins}
\usepackage{tabularx}
\usepackage{multirow}

\usepackage[inline]{enumitem} 

\definecolor{blue}{HTML}{1F77B4}
\definecolor{orange}{HTML}{FF7F0E}
\definecolor{green}{HTML}{2CA02C}

\pgfplotsset{compat=1.14}

\begin{document}
\title{Down the bot hole:
actionable insights from a 1-year analysis of bots activity on Twitter}
%
\date{}
%
%
\author{Luca Luceri, Felipe Cardoso, and Silvia Giordano \\
\small University of Applied Sciences and Arts of Southern Switzerland (SUPSI)\\ 
\footnotesize luca.luceri@supsi.ch, felipe.cardoso@supsi.ch, silvia.giordano@supsi.ch}


%
\maketitle              
\begin{abstract}

Nowadays, social media represent persuasive tools that have been progressively weaponized to affect people's beliefs, spread manipulative narratives, and sow conflicts along divergent factions. Software-controlled accounts (i.e., bots) are one of the main actors associated with manipulation campaigns, especially in the political context. Uncovering the strategies behind bots' activities is of paramount importance to detect and curb such campaigns. In this paper, we present a long term (one year) analysis of bots activity on Twitter in the run-up to the 2018 U.S. Midterm Elections. We identify different classes of accounts based on their nature (bot vs. human) and engagement within the online discussion and we observe that hyperactive bots played a pivotal role in the dissemination of conspiratorial narratives, while dominating the political debate since the year before the election. 
Our analysis, on the horizon of the upcoming U.S. 2020 Presidential Election, reveals both alarming findings of humans’ susceptibility to bots and actionable insights that can contribute to curbing coordinated campaigns. 

\end{abstract}
\section*{Introduction}

False narratives, fake accounts, low-credibility news sources, state-sponsored operators, and so on and so forth: The online ecosystem landscape appears loaded with threats and malicious entities disposed to undermine the integrity of social media discussions. Among those, bots (i.e., automated and software-controlled accounts) \cite{ferrara2016rise,cresci2020decade} and trolls (i.e., human operators often state-sponsored) \cite{zannettou2019disinformation,luceri2020detecting,badawy2019characterizing} have been recognized as the main responsible actors of manipulation and misinformation operations in diverse contexts \cite{ferrara2015manipulation}, ranging from finance \cite{nizzoli2020charting,cresci2018fake,tardelli2020characterizing} to public health \cite{ferrara2020types,yang2020prevalence}, in which the rise of \textit{infodemics} (i.e., the widespread diffusion of unverified information and conspiracy theories) during the Covid-19 outbreak represents the latest milestone of the misinformation age \cite{zarocostas2020fight}.
Moreover, the abusive behavior of these malicious actors received enormous resonance in the political domain \cite{luceri2019red,bessi2016social,metaxas2012social,howard2016bots,allcott2017social,Badawy2018,gerber2016does,zannettou2019let,luceri2019evolution}, where 
the abuse of social platforms has put under threat the effective fulfillment of the democratic process, other than creating worldwide concerns for the integrity of voting events \cite{stella2019influence,broniatowski2018weaponized,stewart2018examining,badawy2019characterizing,ratkiewicz2011detecting,ferrara2017disinformation,howard2017junk,shu2017fake,vosoughi2018spread,stella2018bots,bovet2019influence, Grinberg2019, scheufele2019science, ruck2019internet}.

In such a context, recent findings showed that most of the political messages shared on Twitter are published by a small set of hyperactive accounts \cite{yang2020twitter,hughes2019national}.
Yang \textit{et al.} \cite{yang2020twitter} observed that hyperactive users are more likely to publish low-credibility narratives with respect to ordinary users and they tend to exhibit suspicious behaviors, often linked to automated and fake accounts. In this regard, Ferrara \textit{et al.} \cite{ferrara2020election} recognized how a set of \textit{conspiracy} bots pushed a relevant portion of low-credibility information within the 2020 U.S. Presidential election debate on Twitter.
From this perspective, it follows that, by flooding social platforms with content of questionable accuracy, hyperactive accounts can both manipulate organic users, by affecting and influencing their opinions, but also the platform mechanisms and its engagement metrics, e.g., trending topics and feed ranking algorithms \cite{yang2020twitter}.
Such vulnerabilities, along with the relentless presence of inauthentic entities, highlight the need for intervention to prevent the distortion of online discussions and the manipulation of public opinion.

Therefore, developing computational tools to detect deceptive and orchestrated campaigns is the main goal of the agenda initiated by the research community to purge online platforms. However, this represents a challenging task \cite{ferrara2016detection,chen2018unsupervised,varol2017early}.
Nevertheless, in the last few years, researchers offered
several approaches to identify bots \cite{varol2017online,yang2019arming,cresci2019better,mazza2019rtbust,sayyadiharikandeh2020detection,cresci2020emergent,chavoshi2016debot,subrahmanian2016darpa,chen2018unsupervised},
while solutions for unveiling the activity of trolls have been recently proposed \cite{luceri2020detecting,addawood2019linguistic}.
Other studies focused on the detection of collective and inauthentic behaviors of malicious accounts to uncover coordinated campaigns \cite{sharma2020identifying, nizzoli2020charting,pacheco2020uncovering} and suspicious content diffusion \cite{hui2020botslayer,giglietto2020coordinated,zannettou2019characterizing}.
However, social media abuse persists and the online ecosystem still presents a mix of organic and malicious users \cite{luceri2019evolution,im2019still}, where the former class still demonstrates a moderate capability to identify the latter \cite{yan2020asymmetrical}. This also calls for 
a clear understanding of users' susceptibility to the content shared by malicious accounts and the interplay with them.

\paragraph{Research Questions and Contributions}
In light of these considerations, more research is needed to uncover the strategies behind the activity of malicious actors on social media for curbing the abuse of online platforms, as well as to investigate the downstream effects of users’ exposure to and interaction with malicious accounts for appraising the impact of their deceptive activity. 

Along these research directions, in this paper, we inspect the activity of bot and human accounts on Twitter during the year approaching the 2018 U.S. Midterm elections, which were held on November 6, 2018.
To the best of our knowledge, this work represents the first attempt to study users, and in particular bots, temporal activity over an extended period of time (one year).
The aim is to explore the strategies developed to both inject automated accounts into the Midterm debate and program their sharing activities while avoiding detection.
We focus on the sharing activities that Twitter users can employ to create content (i.e., \textit{original tweets}), re-share others' tweets (i.e., \textit{retweets}), and respond to others’ tweets (i.e., \textit{reply}).
We investigate how and to what extent bots used such \textit{digital weapons}  \cite{cardosodigital} over time to identify cues that can empower the identification of bot accounts and, accordingly, contribute to the detection of orchestrated campaigns.

More specifically, this paper aims at answering the following Research Questions (RQs):
\begin{itemize}
    \item \textbf{RQ1: How did bots perform their sharing activities over time?}
    We explore the temporal dynamics of bots operations by examining the volume of published content during the year and, in particular, as the election approached.  
    \item \textbf{RQ2: When did bots enter into the Midterm debate over the year?} The rationale is to understand how bots were strategically injected into the online conversation to perform their deceptive activity while avoiding detection.
    \item \textbf{RQ3: Did hyperactive accounts play a pivotal role in the broadcasting of political messages?} We aim to investigate the nature of hyperactive accounts, monitor their appearance in the Midterm discussion, and shed light on their activity. 
    
    \item \textbf{RQ4: How did bot accounts engage and interact with human users?}
    We explore the interactions, in terms of retweets and replies, between humans and bots,  and we measure the centrality within the online discussion of both classes of accounts.
 
\end{itemize}

To respond to these RQs, in this work, we attain the following findings:
\begin{itemize}
    \item[\textbf{RQ1}:] We found that bots followed similar temporal patterns of humans in every sharing activity over the whole observation period, which suggests that bots strategically attempted to mimic human operations since the beginning of the Midterm debate. We observed that bots flooded Twitter with a disproportionate number of retweets, but an increasing number of bots aimed at creating original content entered into the conversation as the election approached.  
    \item[\textbf{RQ2}:] We discovered that a relevant fraction of bots started pushing content related to the Midterm election even one year prior to the election. A constant number of new bot accounts progressively entered into the Midterm debate every week of the year, indicating a cautious strategy to infiltrate bots within the online discussion to avoid their detection. We recognized that another significant fraction of bots appeared the month before the election to operate and supposedly interfere into the debate. 
    \item[\textbf{RQ3}:]  We identified 83k hyperactive accounts (9 percent of the users in our dataset), which were responsible for the creation of more than 70M political messages (72 percent of the collected tweets). We recognized their pivotal role in the broadcasting of conspiratorial narratives and we observed that they exhibited a higher degree of automated behavior if compared to ordinary users. Interestingly, we noticed that a small group of hyperactive bots (19k accounts) were involved in the Midterm discussion since one year before the election, generating a gargantuan volume of political messages (almost 30.6M tweets), and we found that most of these accounts have also been operating in the upcoming 2020 U.S. Presidential election debate. 
    
    \item[\textbf{RQ4}:] We measured to what extent humans relied on and interacted with the content shared by bot accounts. We recognized that human users significantly connected to bots by re-sharing their content, while less interplay is noted in terms of replies. Alarmingly,
    we observed that one-third of humans' retweets is a re-share of the content originated by bot accounts.
    We examined users' embeddedness within the social network and found that bots, and in particular the hyperactive and less recent ones, populate the most central and influential area of the social network.
\end{itemize}

\section*{Data and Methodology}
In this Section, we present the Twitter data collected to perform our analysis and the methodologies followed to gather the set of data and detect bot accounts. Then, we detail the proposed approach to distinguish classes of accounts based on their engagement in the Midterm debate. Finally, we describe the technique used to measure the centrality of the accounts within
the Twitter conversation.

\paragraph{Data collection and bot detection}

To perform our analysis, we leverage the dataset gathered by the researchers of the George Washington University \cite{DVN/AEZPLU_2019}. 
The authors collected Twitter messages (i.e., tweets) for one year (from December 2017 to December 2018) by using Midterm elections-related keywords and hashtags.
Based on Twitter’s developer policy, the authors \cite{DVN/AEZPLU_2019} were allowed to publicly share only the tweet IDs of the collected tweets. 
Given the released tweet IDs, a process referred to as \textit{hydration} was necessary to recover the complete tweet information. The hydration process uses the \textit{Twitter API} to obtain the entire tweet object from the corresponding tweet ID. 
Only 74\% of the released 171,248,476 tweet IDs were still available online at the time we performed the hydration process (December 2019) and, as a consequence, we were able to gather only this subset of tweets.

We then considered narrowing the analysis to the tweets published by a set of accounts studied in our previous works \cite{luceri2019evolution,deb2019bots} to enable comparison and validate previous findings.
Moreover, from the set of 997k users analyzed in \cite{luceri2019evolution,deb2019bots}, we found 943k accounts in the collected dataset that
were responsible for the creation of the majority of the collected tweets. Specifically, more than 98M tweets (over the 126M collected) were shared by the set of 943k accounts.
It should also be noted that a consistent subset of such users (62 percent), which is responsible for almost 80 percent of the collected tweets, is also active at the time of this writing (mid-October 2020) in the debate related to the U.S. 2020 Presidential election \cite{chen2020election2020}. The interested reader can find further details in the \textit{Supplement}.

Given that the set of 943k accounts was analyzed in previous efforts, we rely on the account classification (bot vs. human) that we performed and validated in \cite{luceri2019evolution,deb2019bots}. 
The accounts classification was performed by means of Botometer\footnote{\url{https://botometer.iuni.iu.edu/}}, a machine learning-based tool developed by Indiana University \cite{davis2016botornot,varol2017online}.
Botometer outputs a bot score: the lower the score, the higher the
probability that the user is human. 
Besides using the bot score to classify the accounts, we further analyzed the status of such accounts, i.e., active, suspended, or deleted (not found) by leveraging the Twitter API.
In \Cref{table:dataset_full_stats}, we show the statistics of the final dataset under analysis.

\begin{table}[h!]
\small
\vspace{+.25cm}
    \caption{Accounts and tweets statistics}
    \vspace{+.1cm}
    \centering
    \begin{tabular}{l|l|ll|ll}
    
    {} &       \textbf{All} &      \textbf{Bots} &    \textbf{Humans} & \textbf{Suspended} & \textbf{Not found} \\
    \hline
    \textbf{Accounts        } &   943,126 &   184,545 &   758,581 &    38,164 &    30,688 \\
    \textbf{Accounts \%} &       &    19.57\% &    80.43\% &     4.05\% &     3.25\% \\
    \hline
    \textbf{Tweets          } &  98,123,612 &  42,386,269 &  55,737,343 &  9,645,522 &  2,828,963 \\
    \textbf{Tweets \%  } &       &     43.2\% &     56.8\% &     9.83\% &     2.88\% \\
    \end{tabular}
    \vspace{+.25cm}
    \label{table:dataset_full_stats}
\end{table}

Specifically,
\Cref{table:dataset_full_stats} details the number of accounts and shared tweets for each class of users.
The percentage of bots (around 20 percent) and human accounts (around 80 percent) is in line with previous works \cite{luceri2019evolution,deb2019bots}. The majority of users are still active (around 93 percent) at the time of this writing (mid-October 2020), while a small fraction deleted their account (around 3 percent) or have been suspended by Twitter (around 4 percent). It is out of the scope of this paper to characterize the suspended accounts. However, an evaluation about the correlation between account suspension and the likelihood of automation is provided to the interested reader in the
\textit{Supplement}.


    

\paragraph{Engagement-based accounts classification }
\label{account_classification}

To better understand how both human and bot accounts acted over time, we analyze their behavior according to their engagement within the Midterm debate. 
In particular, we consider the
frequency and duration of their activity during the observation period by identifying, for each account, the following parameters:
\begin{itemize}
    \item \textit{first tweet}: the first time the account shared a tweet within our dataset;
    \item \textit{last tweet}: the last time the account shared a tweet within our dataset;
    \item \textit{tweets count}: the number of tweets shared by the account within our dataset;
    \item \textit{active days}: the number of days between the last and first tweet;
    \item \textit{daily tweet frequency}: the average number of tweets shared by the account during the \textit{active days} time window, computed as:    \begin{equation*}
        \textit{daily tweets frequency} = \frac{\textit{tweets count}}{ \textit{active days}}
    \end{equation*}
\end{itemize}

\begin{figure}[h!]
    \centering
    \includegraphics[width=\linewidth]{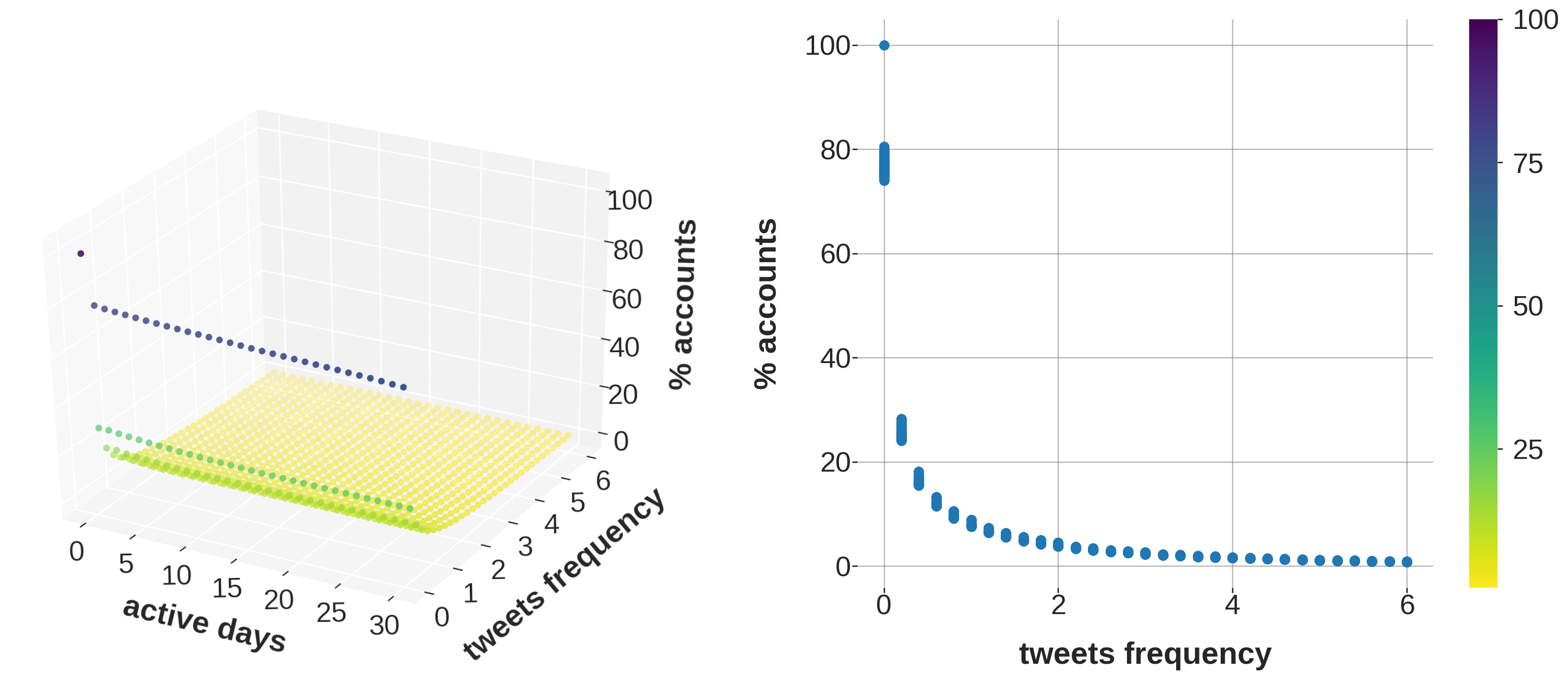}
    \caption{Percentage of accounts that persist to the filtering based on the \textit{active days} and \textit{tweet frequency} parameters}
    \label{fig:data_filter_3d_accounts}
\end{figure}

Based on the \textit{active days} and \textit{daily tweets frequency} parameters, we propose to classify the accounts into two categories: \textit{hyperactive}
and
\textit{ordinary} accounts. 
To distinguish these two classes of accounts, we need to
identify a threshold for each considered parameter that allows to opportunely separate the hyperactive users from the ordinary ones.
More specifically, we observe
how ordinary (resp. hyperactive) accounts are filtered out (resp. selected) from the whole set of accounts by varying these thresholds. In \Cref{fig:data_filter_3d_accounts}, we show how the percentage of retained accounts decreases as the filtering parameters (i.e., active days and tweets frequency) increase.

In particular, in the left panel of \Cref{fig:data_filter_3d_accounts}, we show the percentage of accounts that persist to such a filtering as a function of the filtering parameters.
In the right panel of \Cref{fig:data_filter_3d_accounts}, we only focus on the two-dimensional relation between the percentage of accounts and the tweet frequency parameter. Every dot in the figures indicates a combination of the filtering parameters, while its color represents the percentage of accounts that persist to the filtering.
Two facts are worth noting. First, the active days parameter does not significantly affect the percentage of filtered accounts, except for the first gap (from 100 percent to 80 percent of the accounts), where the accounts active only one day are filtered out. It can be noticed that, at a fixed tweet frequency, the percentage of filtered accounts does not vary by increasing the active days parameter (dots appear to be aligned). Second, the tweet frequency parameter highly affects the percentage of filtered accounts: as this parameter increases, the percentage of accounts decreases. The most noticeable gap can be appreciated when the tweet frequency parameter value equals 0.2 tweets/day, which results in a filtering of almost 70 percent of the accounts. 

\begin{figure}[h!]
    \centering
    \includegraphics[width=\linewidth]{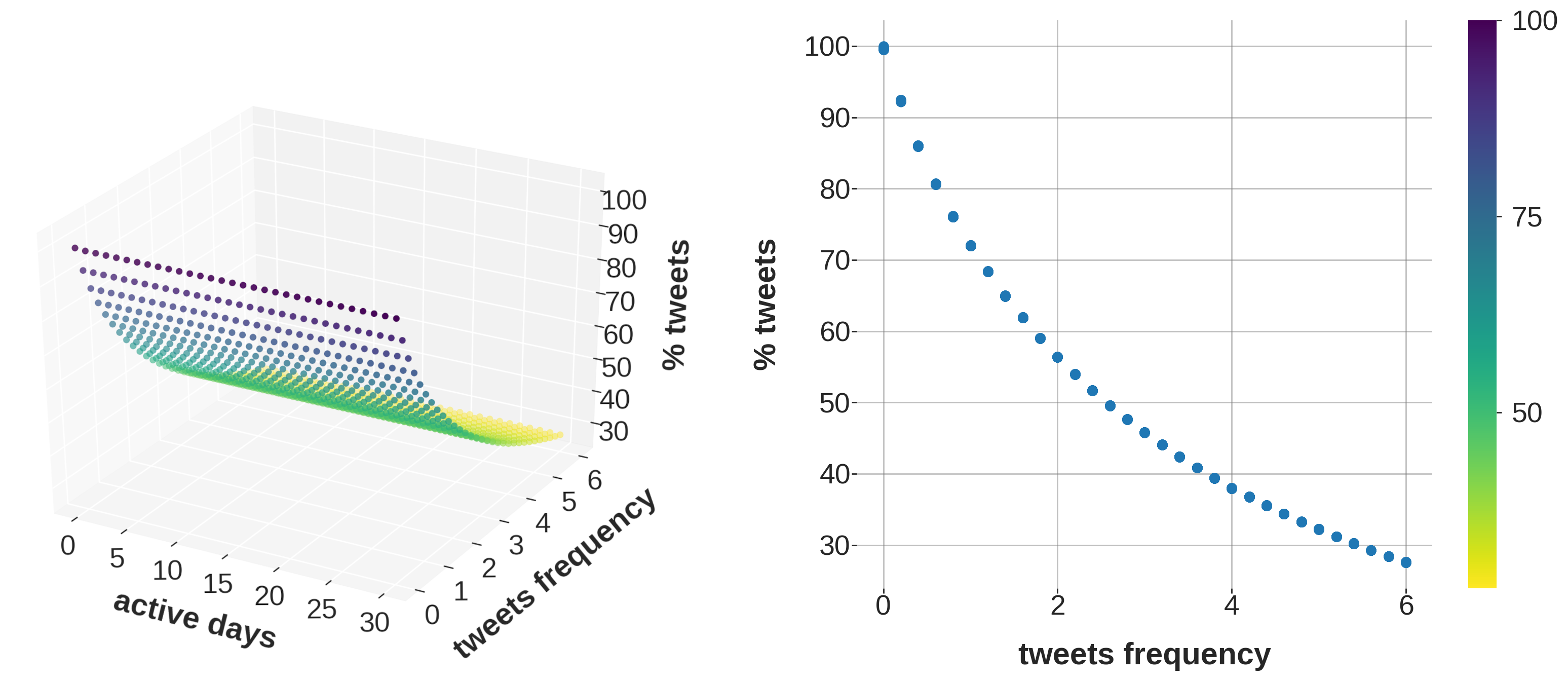}
    \caption{Percentage of tweets that persist to the accounts filtering based on the \textit{active days} and \textit{tweet frequency} parameters}
    \label{fig:data_filter_3d_tweets}
    \end{figure}

In \Cref{fig:data_filter_3d_tweets}, the same evaluation is performed by considering the percentage of tweets that persist to the accounts  filtering (based on the same set of parameters) and similar considerations can be done.

Indeed, also in \Cref{fig:data_filter_3d_tweets}, the parameter that impacts most of the percentage of filtered tweets is the tweet frequency. 
However, in \Cref{fig:data_filter_3d_tweets}, the percentage of tweets decreases more gradually by increasing the tweet frequency parameter if compared to the percentage of accounts reduction (\Cref{fig:data_filter_3d_accounts}). Interestingly, with a tweet frequency of 0.2 tweets/day, more than 70 percent of accounts were filtered out but more than 90 percent of tweets were retained, suggesting that a small percentage of accounts is responsible for a huge amount of tweets, which is consistent with previous findings \cite{hughes2019national}. Given the above observations, we classify as \textit{hyperactive} the accounts that satisfy both the following conditions:
\begin{itemize}
    \item $\textit{active days} \geq 1$ 
    \item $\textit{daily tweets frequency} \geq 1$
\end{itemize}

We consider these parameters as (i) for the \textit{active days} parameter the most noticeable variation in the percentage of filtered accounts can be appreciated when this parameter is greater than 0, and (ii) a \textit{tweet frequency} of 1 corresponds to the knee of the curve in the account distribution (\Cref{fig:data_filter_3d_accounts}), which can be viewed as a conservative choice to discriminate the two classes of accounts (hyperactive vs. ordinary), as no significant difference would occur in the classification by further increasing this threshold, i.e., the percentage of accounts does not significantly vary by increasing the tweet frequency parameter.

\begin{figure}[h!]
    
        \centering
        \includegraphics[width=0.7\linewidth]{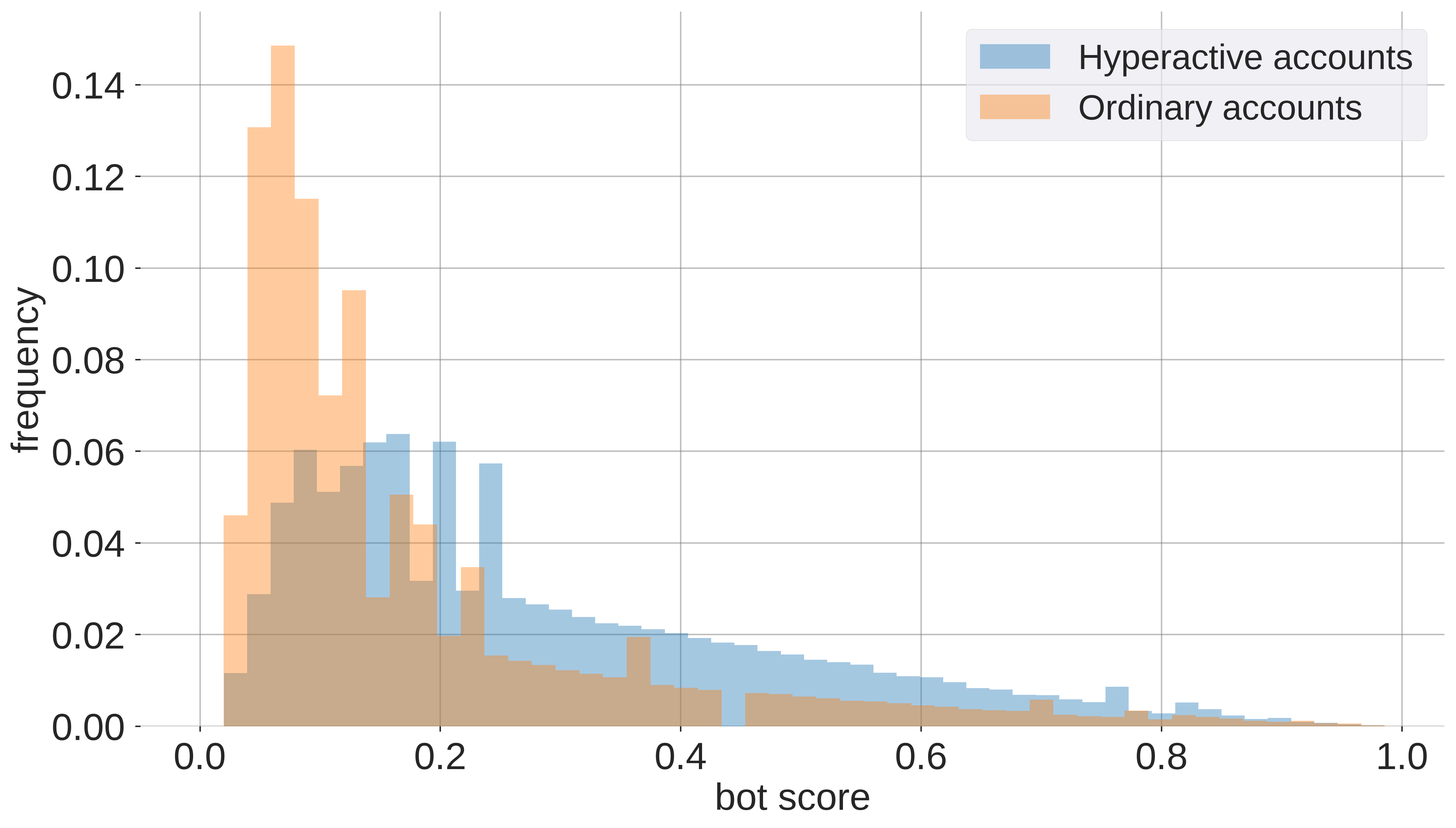}        

\caption{Bot score distributions comparison between hyperactive and ordinary accounts}
\label{fig:bot_scores_kde_active_filter_comp}
\end{figure}

According to these filtering parameters, \textit{hyperactive} accounts represent less than 9 percent of the accounts in our dataset and cover about 70 percent of the collected tweets. 
It is worth noting that bots cover about 38 percent of the \textit{hyperactive} accounts, which represents a noticeable increase of automated accounts with respect to the percentage related to the full set of accounts (80 percent humans vs. 20 percent bots), other than indicating a robust presence of bots within the online conversation. 
This is in line with previous findings \cite{yang2020twitter} and is further confirmed in
\Cref{fig:bot_scores_kde_active_filter_comp}, which depicts the bot score distribution for hyperactive and ordinary accounts. It can be noticed that 
the two classes of accounts exhibit significantly different distributions, as also demonstrated by a Mann-Whitney rank test ($p$-value $<$ 0.001).
On the one hand, 
most of the probability mass in the distribution related to \textit{ordinary} accounts is in the range [0,0.14], suggesting that the majority of ordinary accounts are humans.
On the other hand, in the distribution related to \textit{hyperactive} accounts, we can notice how these users are more likely to have higher bot scores with respect to ordinary accounts, indicating a more relevant presence of automated (software-controlled) users within the set of hyperactive accounts.

\paragraph{$k$-core decomposition}
To evaluate the centrality (or embeddedness) of the accounts within the Twitter conversation, we employ the $k$-core decomposition technique.
The $k$-core decomposition aims to determine the set of nodes deeply embedded in a graph by recursively pruning nodes with degrees less than $k$. Formally, the $k$-core is a subgraph of the original graph in which
every node has a degree equal to or greater than a given value $k$.
The $k$-core decomposition assigns a core number to each node, which is the largest value $k$ of a $k$-core containing that node. The higher the core number $k$ is, the more embedded the node is in the network.

\section*{Results and Analysis}
\label{rq1:weekly}
In this Section, we present our analysis and corresponding results on the temporal and behavioral dynamics of bot and human activity over the whole year preceding the 2018 U.S. Midterm elections.
We focus on investigating bots sharing activities over time with the objective of understanding how such accounts have been strategically injected in the online discussion related to the Midterm as the election approached.
Then, we consider the distinction between hyperactive and ordinary accounts to identify peculiar and distinctive behaviors between these two classes of accounts by also considering their nature (bot vs. human). Finally, we observe the volume of interactions between bots and humans, and their embeddedness within the social network.

\begin{figure*}[h!]
    \centering
        \begin{subfigure}[t!]{.48\textwidth}
            \centering
            \includegraphics[width=\columnwidth]{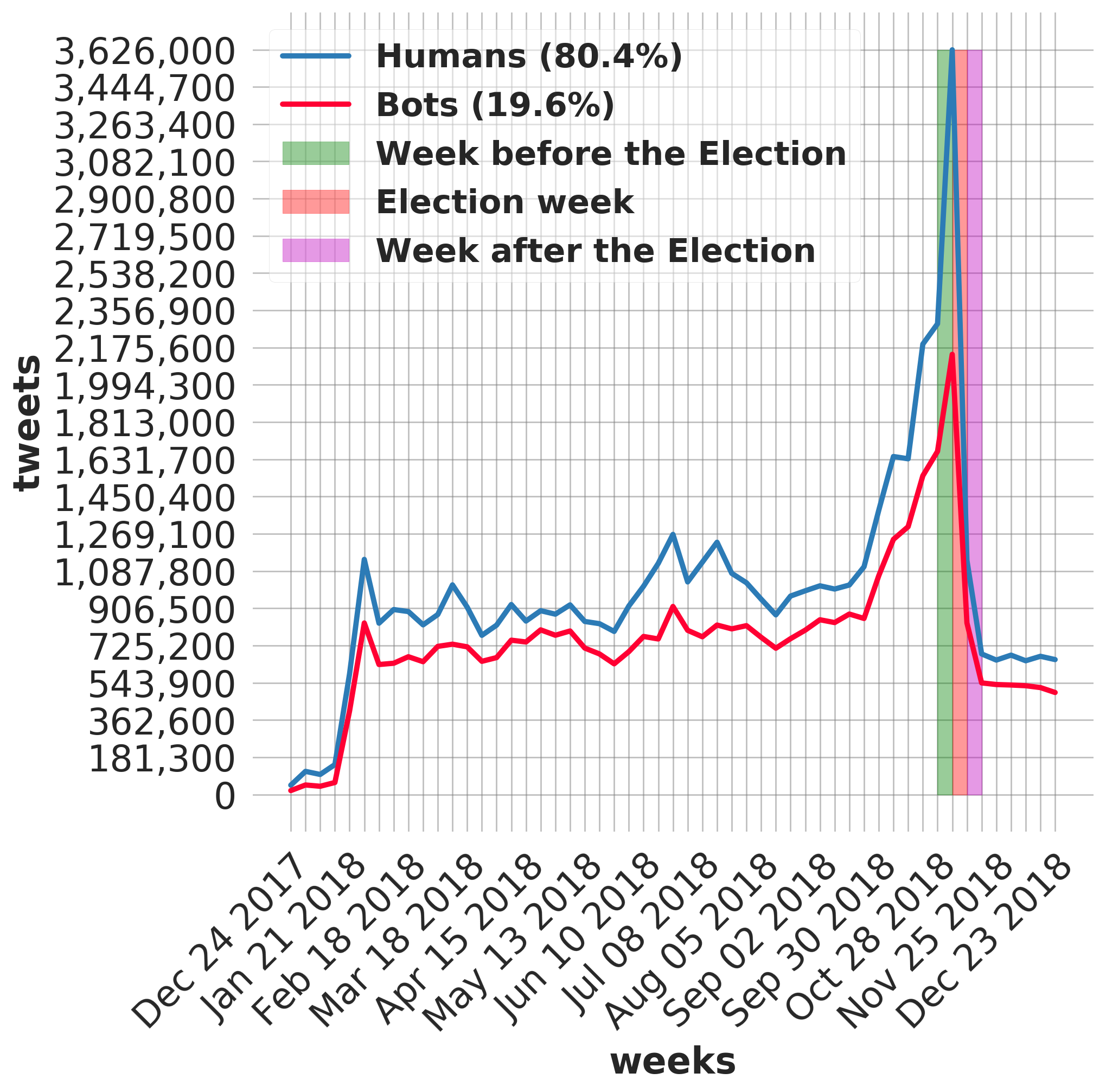}
            \caption{Tweets}
            \label{fig:weekly_activities_full_tweets}
        \end{subfigure}%
        \vspace{0.4pt}
        \begin{subfigure}[t!]{.48\textwidth}
            \centering
            \includegraphics[width=\columnwidth]{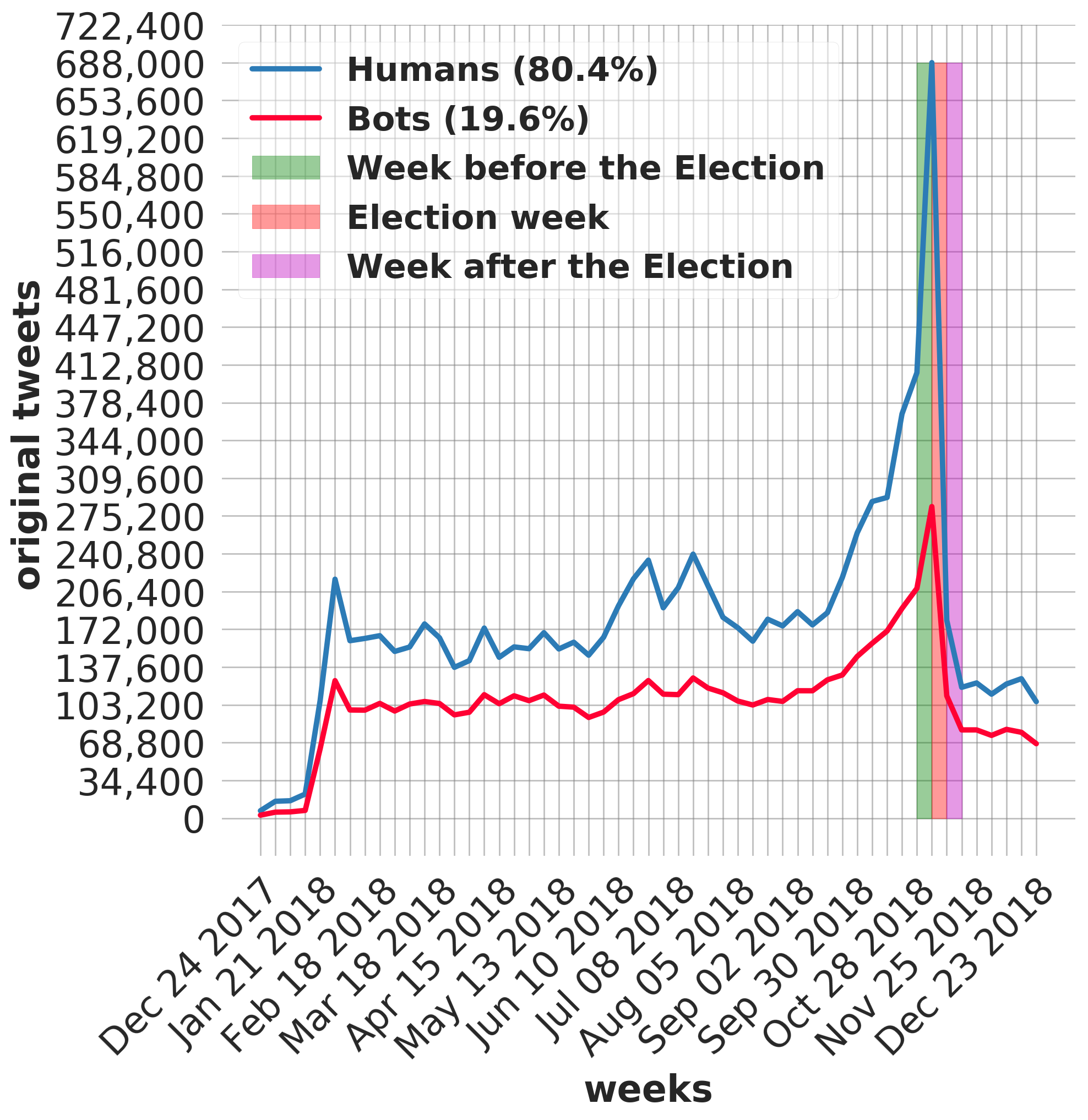}
            \caption{Original tweets}
            \label{fig:weekly_activities_full_original_tweets}    
        \end{subfigure}\\
        \hspace{0.4pt}
        \begin{subfigure}[t!]{.48\textwidth}
            \centering
            \includegraphics[width=\columnwidth]{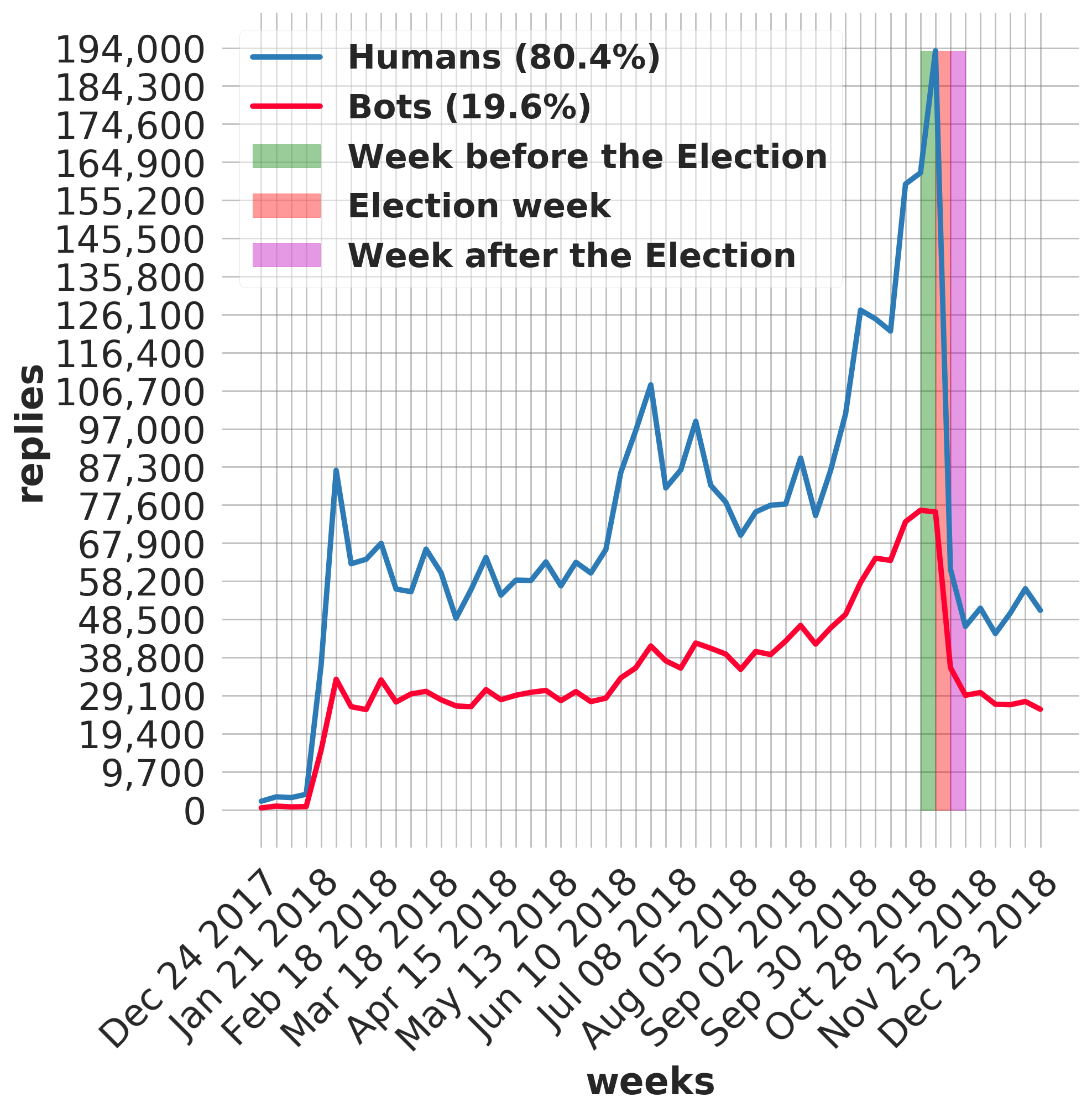}
            \caption{Replies}
            \label{fig:weekly_activities_full_replies}
        \end{subfigure}%
        \vspace{0.4pt}
        \begin{subfigure}[t!]{.48\textwidth}
            \centering
            \includegraphics[width=\columnwidth]{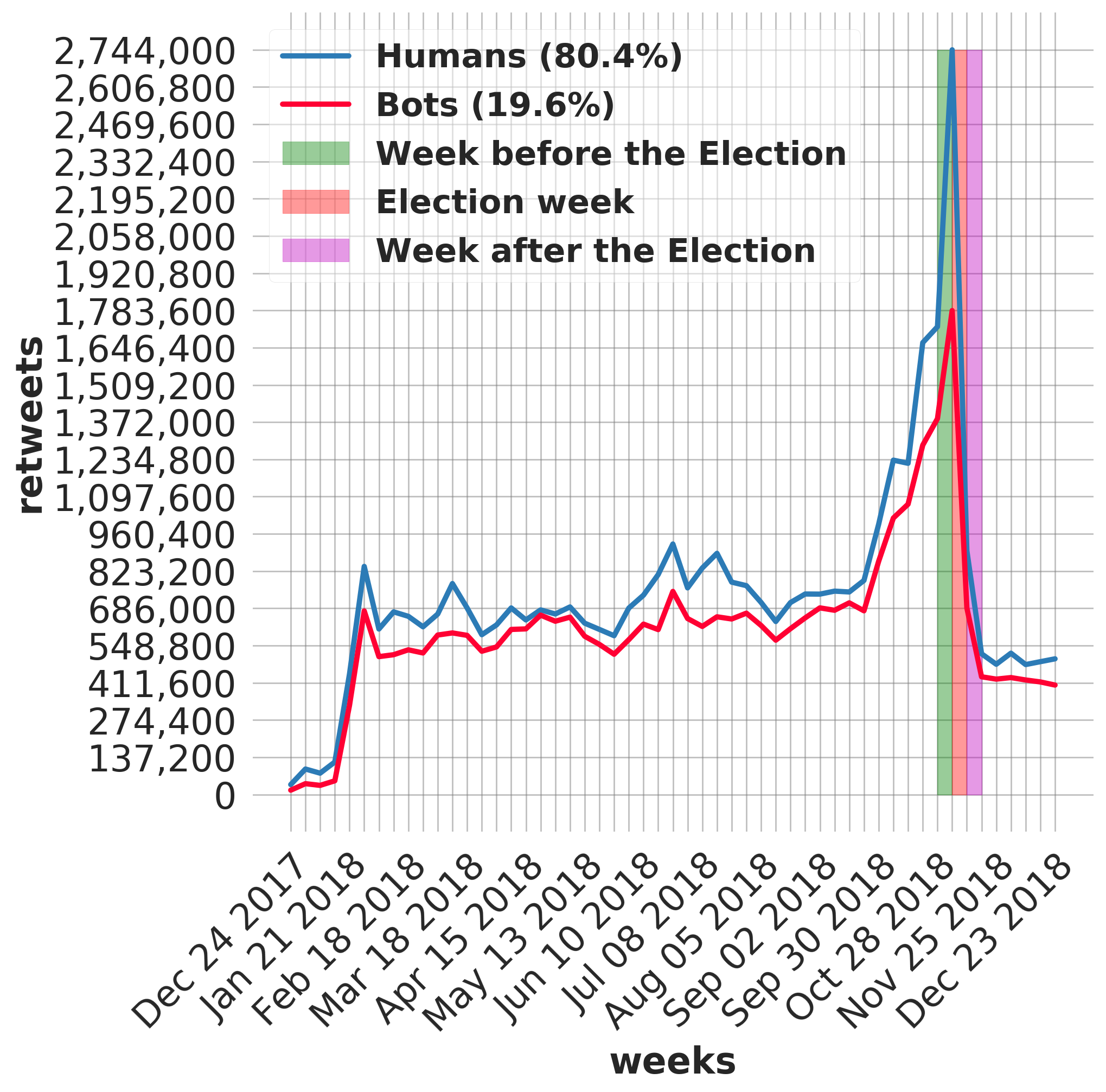}
            \caption{Retweets}
            \label{fig:weekly_activities_full_retweets}    
        \end{subfigure}\\
    \caption{Sharing activities over time of bot and human accounts}
    \label{fig:weekly_activities_full}
\end{figure*}

\paragraph{RQ1: Bots activity over the Midterm year}
In \Cref{fig:weekly_activities_full_tweets}, we show the number of tweets shared by human and bot accounts per each week of the observation period.
It could be noticed that the human population shared more tweets than bots over the whole year.
However, this could be expected given that, as shown in Table \ref{table:dataset_full_stats}, human accounts represent the largest portion (80 percent) of the accounts of the collected dataset and shared more content (about 57 percent of the tweets) with respect to the bot population (about 20 percent of the accounts and 43 percent of the tweets).
As a consequence, bots were more active (in terms of tweets generated per account) than humans during the whole period.
On average, a bot shared three times the number of tweets published by a human user (218 vs. 70 tweets/account), which is consistent with bots' purpose of flooding online platforms with a high volume of tweets \cite{bessi2016social}.
Interestingly, bots followed a similar temporal pattern of human users over the whole year, suggesting that bots strategically attempted to mimic human sharing activity since the beginning of the online conversation related to the Midterm elections.
As expected, the content shared by both classes of accounts increased as the election approached, and a spike of activity is noticeable during the election week. 

To better examine how bots performed their activity over time, we disentangle the shared messages in \textit{original tweets}, \textit{retweets}, and \textit{replies}.
In \Cref{table:dataset_full_detailed_group_stats}, we show the number of shared messages for each category of tweet and class of accounts over the observation period.

\begin{table}[h]
\vspace{+.25cm}
    \caption{Sharing activities of bot and human accounts}
    \vspace{+.25cm}
    \centering
    \begin{tabular}{l|l|ll}
    
    {} &      \textbf{Bots} &    \textbf{Humans}\\
    \hline
    \textbf{Accounts        } &      184,545 &   758,581\\
    \hline
    \textbf{Original tweets          } &  5,677,142 (14\%) &  9,635,364 (18\%)\\
    \hline
    \textbf{Retweets          }  &  32,746,675 (81\%) &  39,427,132 (74\%)\\
    \hline
    \textbf{Replies          }&  1,849,625 (5\%) &  3,876,595 (8\%)\\
    \end{tabular}
    \vspace{+.25cm}
    \label{table:dataset_full_detailed_group_stats}
\end{table}

It can be noticed that retweeting is the most used operation for both humans (74 percent of the time) and bots (81 percent of the time). As expected, bots heavily relied on the retweet action as it represents a simpler operation for automated accounts with respect to the creation of an original tweet or a reply, which requires the usage of more sophisticated techniques based on natural language processing and understanding \cite{ferrara2019history}. This is also in line with previous findings \cite{luceri2019evolution}, further confirming that retweets have been employed as bots' favorite digital weapon to resonate messages and create an illusion of consensus within the online community \cite{ferrara2016rise}.
Figures \ref{fig:weekly_activities_full_original_tweets}-c-d portray the number of original tweets, replies, and retweets shared weekly by humans and bots over the year approaching the Midterm elections.
Interestingly, also in the disentangled sharing activities, bots followed a similar temporal pattern of human users over the whole year. Indeed, the number of tweets shared by humans and bots over time positively correlate ($\rho>0.97$, $p$-value $<0.001$) for each kind of activity (i.e., original tweets, retweets, and replies).
Although original tweets and replies require to develop advanced AI techniques on software-controlled accounts, the volume of such activities performed by bots approaches human activities volume (see also Table \ref{table:dataset_full_stats}), especially if we consider the number of messages published per each account.
This suggests that also more sophisticated bots operated along with less advanced \textit{spam bots} \cite{ferrara2019history}. 
Also, similarly to \Cref{fig:weekly_activities_full_tweets}, the content shared by both classes of accounts increased as the election approached, and a spike of activity is noticeable during the election week. Finally, it is worth noting how bots started emulating every kind of human sharing activity since the beginning of the year, further confirming how detecting coordinated campaigns is a challenging task, even when users are monitored over an extended observation window.

\begin{figure*}[h!]
\centering
\begin{subfigure}[t!]{.5\textwidth}
\centering
\includegraphics[width=\columnwidth]{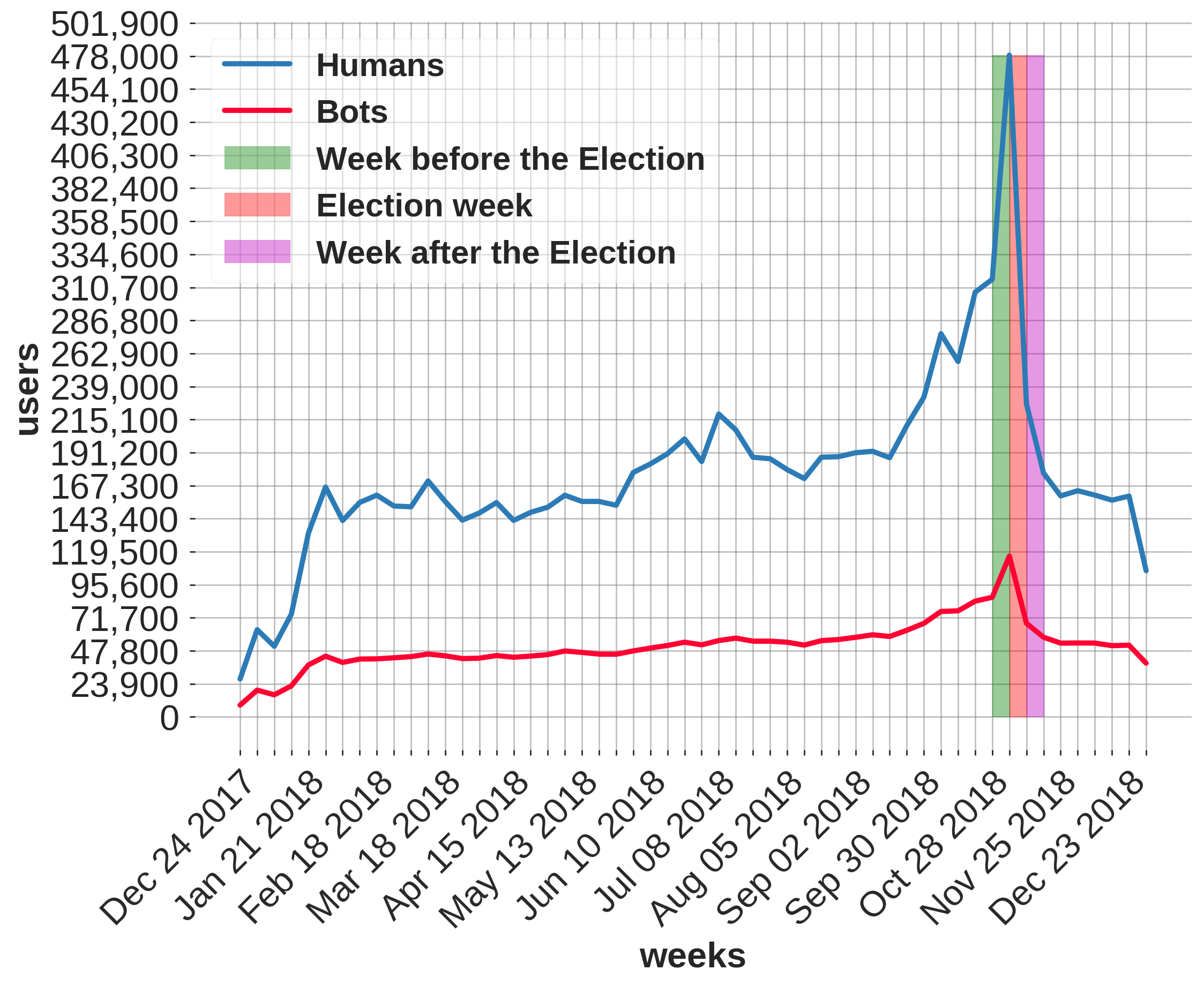}
\caption{Engaged accounts}
\label{fig:weekly_users_full}
\end{subfigure}%
\begin{subfigure}[t!]{.5\textwidth}
\centering
\includegraphics[width=\columnwidth]{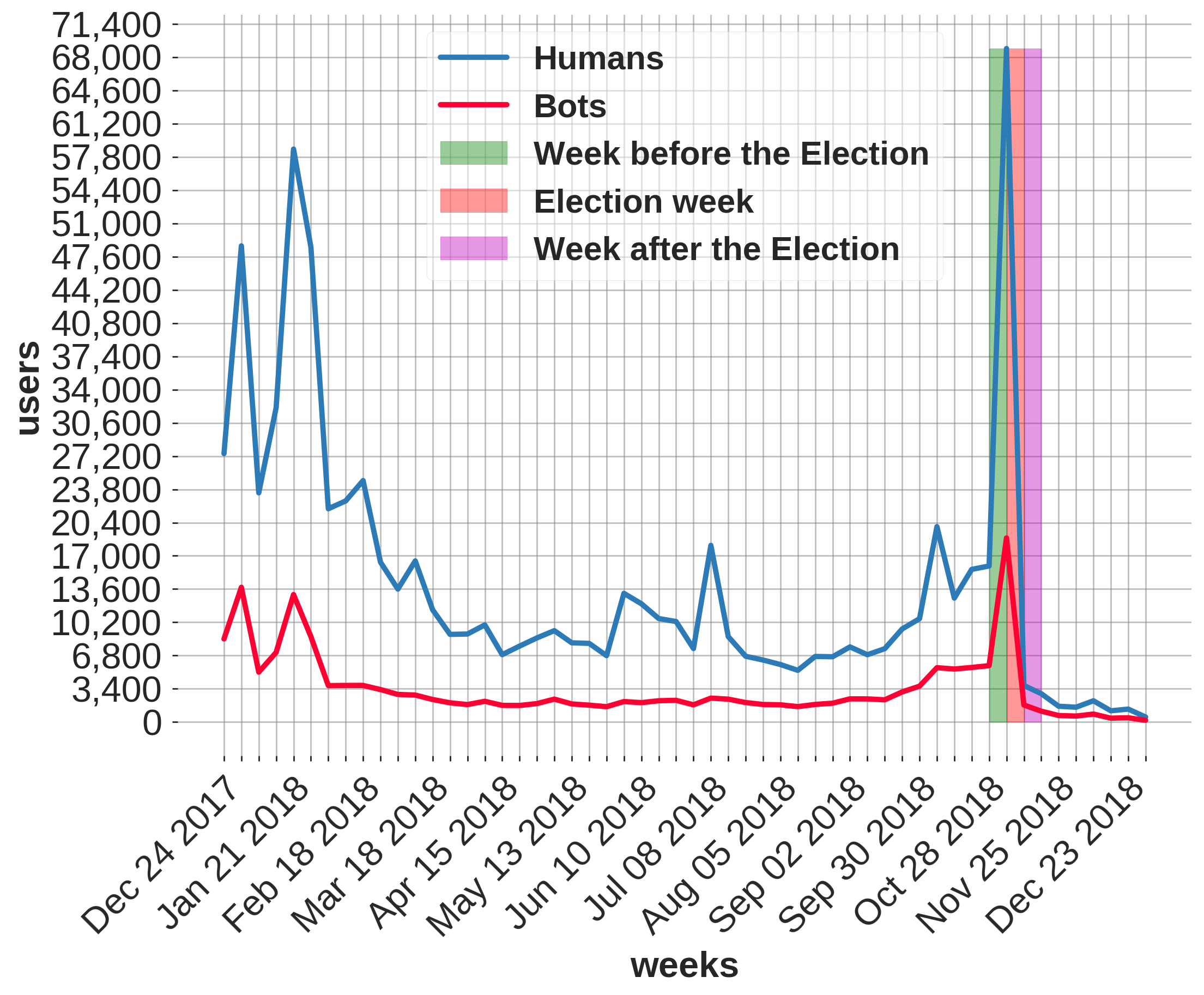}
\caption{New accounts}
\label{fig:weekly_new_users_full}
\end{subfigure}\\
\caption{Engaged and new accounts within the Midterm debate}
\label{fig:test}
\end{figure*}

\paragraph{RQ2: Bots injection within the Midterm debate}
To have a more comprehensive understanding of how both human and bot online activities evolve over time, in \Cref{fig:weekly_users_full}, we depict the number of human and bot accounts engaged in the debate every week of the observation period. In a given week, we denote as \textit{engaged} an account that shared at least one tweet during the week under analysis.

It should be noticed that the number of engaged bots per week is almost steady and slowly increases over time, while human users engagement is less regular. This controlled engagement of bots might suggest a coordinated effort to simultaneously avoid the detection of such automated accounts, while infiltrating them into the political discussion.

To shed light on this finding, we observe the number of new accounts that joined the conversation related to the Midterm elections during the observation window. We define an account as \textit{new} when it participates for the first time in the Midterm debate.
In \Cref{fig:weekly_new_users_full}, we display the number of new accounts for each week of the period under analysis. 
Once again, bot temporal pattern shows a stationary trend, where an almost constant number of new accounts entered into the conversation each week. 
As we mentioned before, this controlled, and supposedly scheduled, activity appears a more cautious approach for injecting bots within the online discussion to allegedly avoid their detection.
Moreover, we notice that a significant number of accounts (34 percent of the accounts within our dataset) joined the Midterm conversation since the end of 2017 and beginning of 2018 (first two weeks of February), as it can be observed from the two peaks in the left side of  \Cref{fig:weekly_new_users_full}. Interestingly, this subset of accounts is responsible for 80 percent of the collected tweets, suggesting that such users play a central role within the Twitter discussion. This holds true both for human (28 percent of the accounts) and bot (6 percent of the accounts) users. 
It is worth, and mostly alarming, noting that 60k of such accounts are bots that have persisted online during the whole year generating around 34.2 million tweets (35 percent of the tweets) with an average rate of 570 tweets per account.
It should be considered that this subset of 60k bots represents a relevant fraction (about 33 percent) of the bot accounts that participated in the Midterm discussion and most of them (44k accounts) are actively involved in the 2020 U.S. Presidential election debate at the time of this writing (mid-October 2020).
Also, we observe that the 260k human accounts that appeared in the early stage of the year (28 percent of the accounts) shared 44.9 million tweets (45 percent of the tweets) with an average rate of 173 tweets per account, which is about 3.3 times lower than bots tweet fabrication rate.     

A similar trend of appearance can be observed in the time around the election day. Specifically, from the first week of October to the week after the election, we observe a peak corresponding to a new set of accounts (11.6 percent of the total users) that joined the conversation. Among such accounts, we recognize 85k human users and 24.6k bots, which were responsible for creating 210k and 158k tweets during the aforementioned period, respectively. It could be noticed that the ratio between the sharing rate of bots (6.4 tweets/account) and humans (2.5 tweets/account) is lower than the one observed with the set of accounts that joined the conversation since the beginning of the year (3.3 vs 2.5), which allegedly indicates a more cautious strategy adopted by bots for sharing content in a relatively short time window (if compared to the whole year).

To characterize the behavior of bot accounts that participated in the Midterm debate, in \Cref{table:old_vs_election_bots}, we compare the sharing activities of those that joined the conversation since the beginning of the year (from now on referred to as \textit{old bots}) with those that appeared during the election period (from now on referred to as \textit{election bots}). It can be noticed how both classes of bots used the retweet as the main sharing activity. However, it can be observed a relevant difference in the propensity of re-sharing and producing original tweets between old and election bots. Indeed, election bots significantly created more original content ($t$-test results: $t$(4,708,912)=118.2, $p<0.001$), while re-sharing significantly fewer posts with respect to old bots ($t$-test results: $t$(28,213,616)=121.1, $p<0.001$), which might suggest the deployment of different strategies between the two classes of bot accounts based on their lifetime within the Midterm debate.

\begin{table}[h]
    \caption{Sharing activities of old and election bots}
    \centering
    \begin{tabular}{l|l|ll}
    
    {} &      \textbf{Old bots} &    \textbf{Election bots}\\
    \hline
    \textbf{Tweets       } &      34,197,174 &   158,708\\
    \hline
    \textbf{Original tweets          } &  13\% &  25\%\\
    \hline
    \textbf{Retweets          }  &  83\% &  70\%\\
    \hline
    \textbf{Replies          }&  4\% &  5\%\\
    \end{tabular}
    \label{table:old_vs_election_bots}
\end{table}

Notice that all the other bots (i.e., the ones injected between the end of February and the end of September) exhibited a hybrid behavior between the two considered classes of accounts,
with a distribution of shared activities in between old and election bots.
This further highlights that an increasing number of bots created with the purpose of sharing original content have been strategically injected in the online conversation as the election approached.

\begin{figure*}[h!]
    \centering
    \captionsetup{justification=centerlast}
    \begin{minipage}[t]{.50\textwidth}
        \centering
        \includegraphics[width=\textwidth]{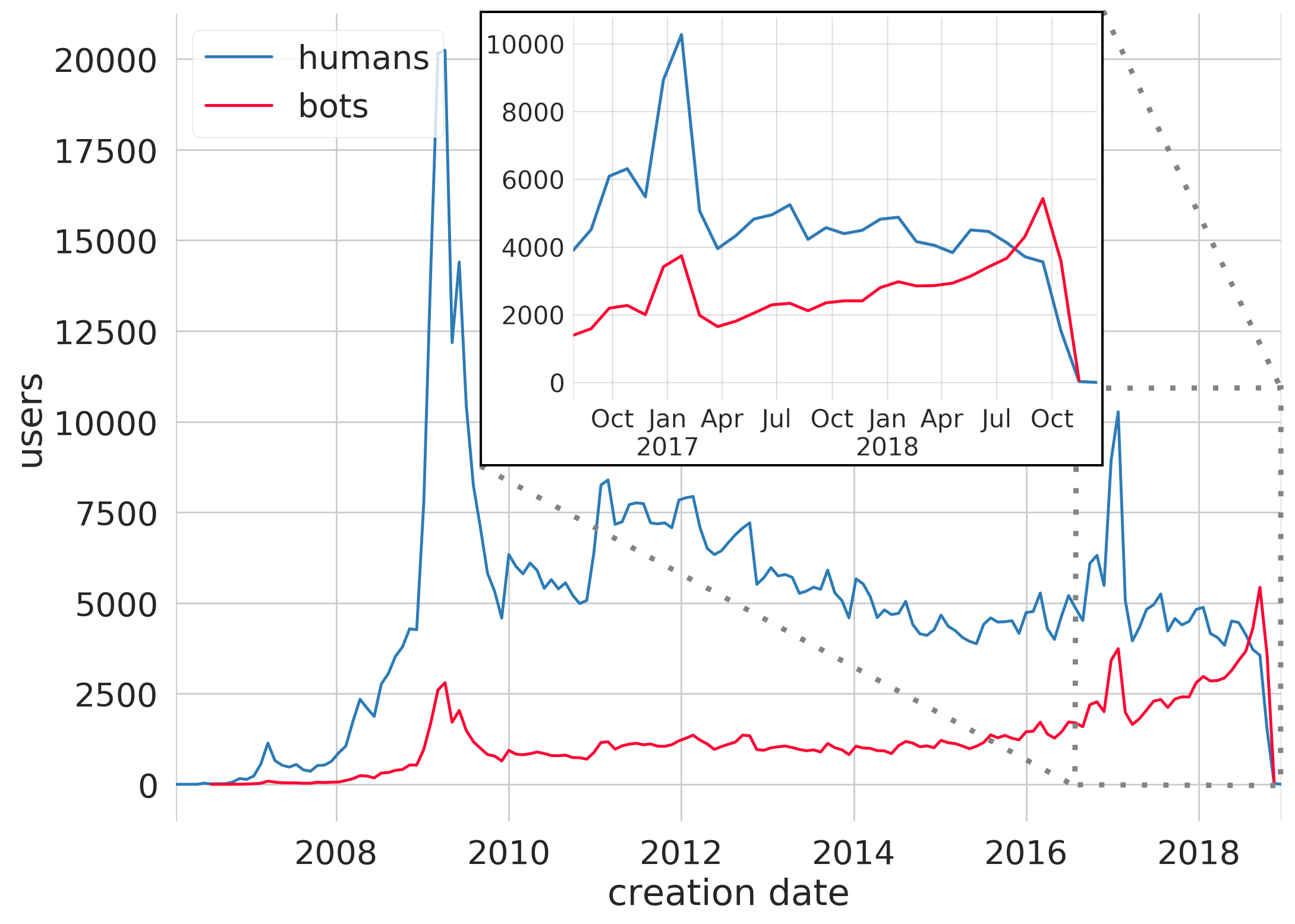}
        \caption{Account creation time}
        \label{fig:creation_time_all}
    \end{minipage}\hfill
    \begin{minipage}[t]{.42\textwidth}
    \centering
    \includegraphics[width=\textwidth]{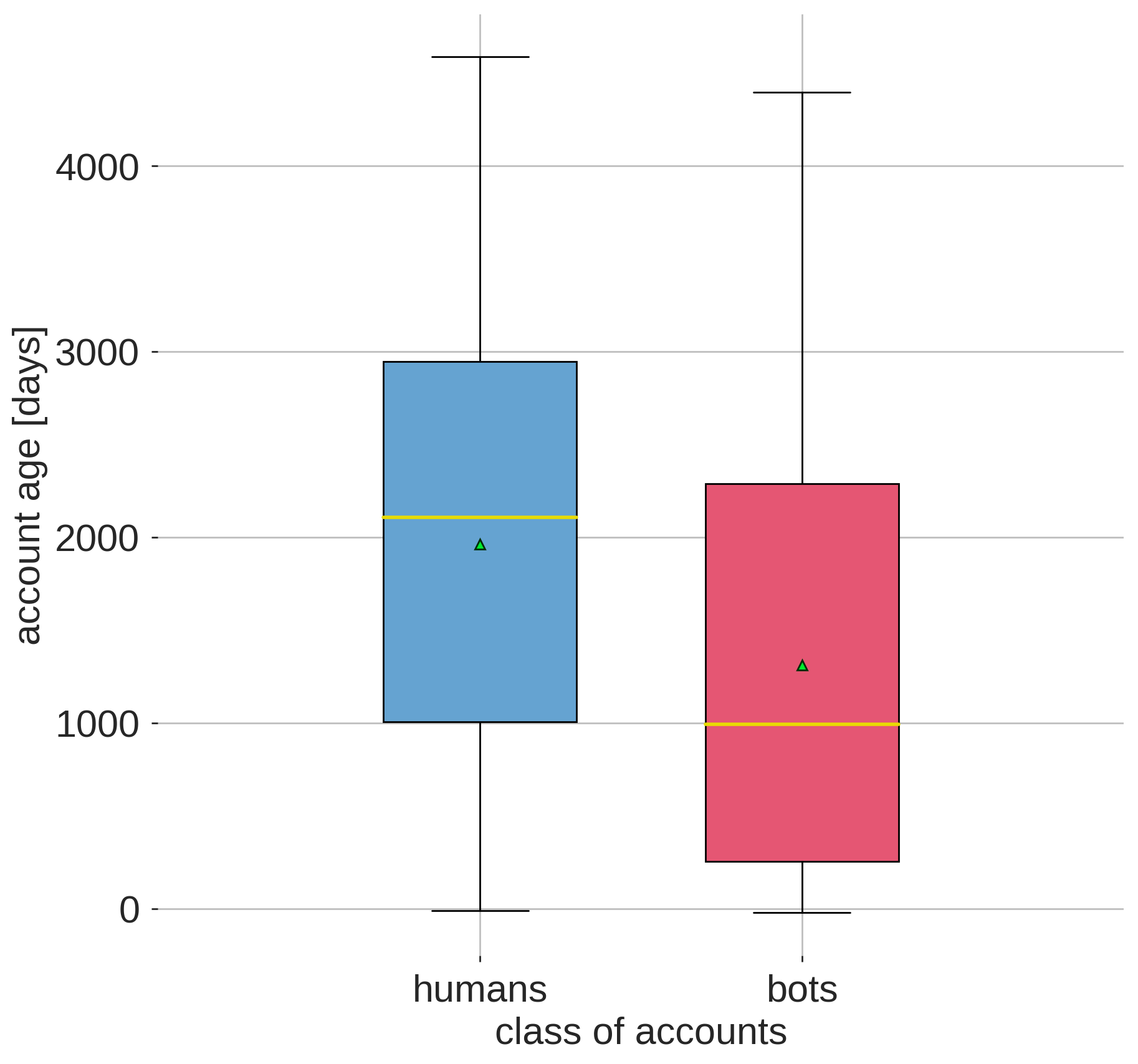}
    \caption{Account age}
    \label{fig:users_delta_activation}
    \end{minipage}\\
    \label{fig:creation_time}
\end{figure*}

To further investigate the nature of the analyzed bot accounts, we examine their creation date (by inspecting tweets metadata). In \Cref{fig:creation_time_all}, we display the date of the creation of bot and human accounts. We observe how the number of created bots within our dataset increases as the election approaches. It stands out how in 2009 (similarly to \cite{yang2020twitter}), 2017, and right before the Midterm elections, an elevated number of bots have been created. It is particularly concerning to note how such bots persisted online over the years, allegedly participating in diverse social and political discussions.
From the superimposed illustration of \Cref{fig:creation_time_all}, it can be appreciated how the bot accounts created from July 2018 to November 2018 (Midterm election month) outnumber the accounts created by humans in the same time window, which highlights that a conspicuous number of bots might have been purposely created for infiltrating into the discourse related to the Midterm elections.
More specifically, about 10 percent of bot accounts were created from July to November 2018, whereas
80 percent of them were created before the beginning of 2018. The remaining fraction (10 percent) was created in the first six months of 2018.

Finally, we consider the account age, computed as the time difference between the date of the first tweet of the account in our dataset and its creation date.
\Cref{fig:users_delta_activation} shows the distribution of the account age, measured in days, for both humans and bots. 
It can be appreciated how the account age for bots (average account age = 1963.8 days) is significantly lower ($p<$0.001) with respect to human accounts (average account age = 1313.3 days). The median account age for bots is around 994 days, whereas for humans is about 2108 days.
This result indicates that bots shared Midterm-related content closer to their creation date if compared to human accounts (similarly to \cite{ferrara2020types}), which further suggests that some bots might have been specifically designed to operate in the Midterm debate.

\paragraph{RQ3: Hyperactive accounts behavior}

Based on the engagement-based accounts classification described in Section \textit{Data and Methodology}, we classify users in \textit{hyperactive} and \textit{ordinary} accounts. 
For what pertains to ordinary accounts,
it should be noticed that, from this set of users, we filtered out the accounts that shared only one tweet during the window of observation, as we want to consider these users distinctly and analyze them in contrast to the hyperactive ones. 
Overall, we observe that 74 percent of the accounts are classified as \textit{ordinary} accounts, 9 percent of the accounts are classified as \textit{hyperactive} accounts, and 17 percent of the accounts shared only one tweet during the whole year (see the \textit{Supplement} for further details).
Ordinary accounts were responsible for about 28 percent of the tweets and the proportion between bots and humans within this set of users is consistent with the full set of accounts. A similar accounts distribution can also be noticed for the users that shared only one tweet, which in turn published a negligible volume of messages.

\textit{Hyperactive} users, while representing a tiny fraction of the accounts (about 83.7k), were responsible for about 72 percent of the collected tweets (about 70.7M messages). 
With 31.8k accounts, bots represent about 38 percent of the hyperactive accounts and published more tweets (38 percent of the collected tweets) than hyperactive humans (62 percent of the hyperactive accounts and 34 percent of the collected tweets) despite the latter set of users outnumbers automated accounts, which further highlights bots prevalence in the online discussion.
Alarmingly, we notice that hyperactive accounts were involved in the vast majority (90 percent) of the 1.6 million tweets related to the \textit{QAnon} conspiracy theory\footnote{\url{https://www.bbc.com/news/53498434}} that we identified within our dataset (see the \textit{Supplement} for more details).
This result supports the finding of Yang, \textit{et al.} \cite{yang2020twitter}, related to hyperactive users' propensity of sharing low-credibility information. Moreover, hyperactive bots dominated the broadcasting of such conspiratorial narratives by pushing 62 percent of the content related to the QAnon theories, similarly to the \textit{conspiracy} bots identified in \cite{ferrara2020election}.  It is, therefore, a cause of concern noting that the vast majority of hyperactive bots (22.5k  accounts) are participating in the 2020 U.S. election debate \cite{chen2020election2020} at the time of this writing.

\begin{figure*}[h!]
\centering
\begin{subfigure}[t!]{.48\textwidth}
\centering
\includegraphics[width=\columnwidth]{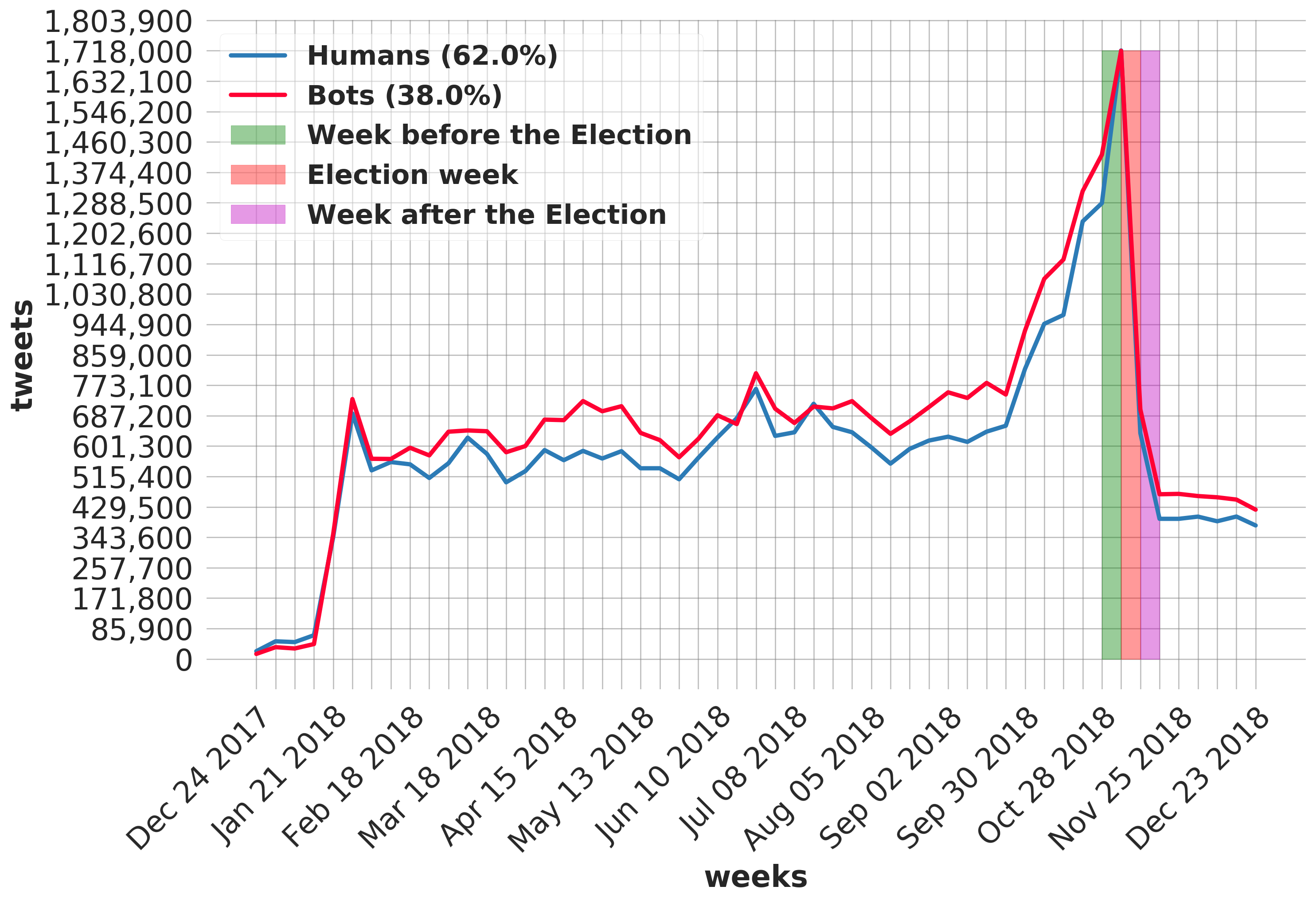}
\caption{Hyperactive accounts}
\label{fig:weekly_tweets_most_active}
\end{subfigure}%
\vspace{0.4pt}
\begin{subfigure}[t!]{.48\textwidth}
\centering
\includegraphics[width=\columnwidth]{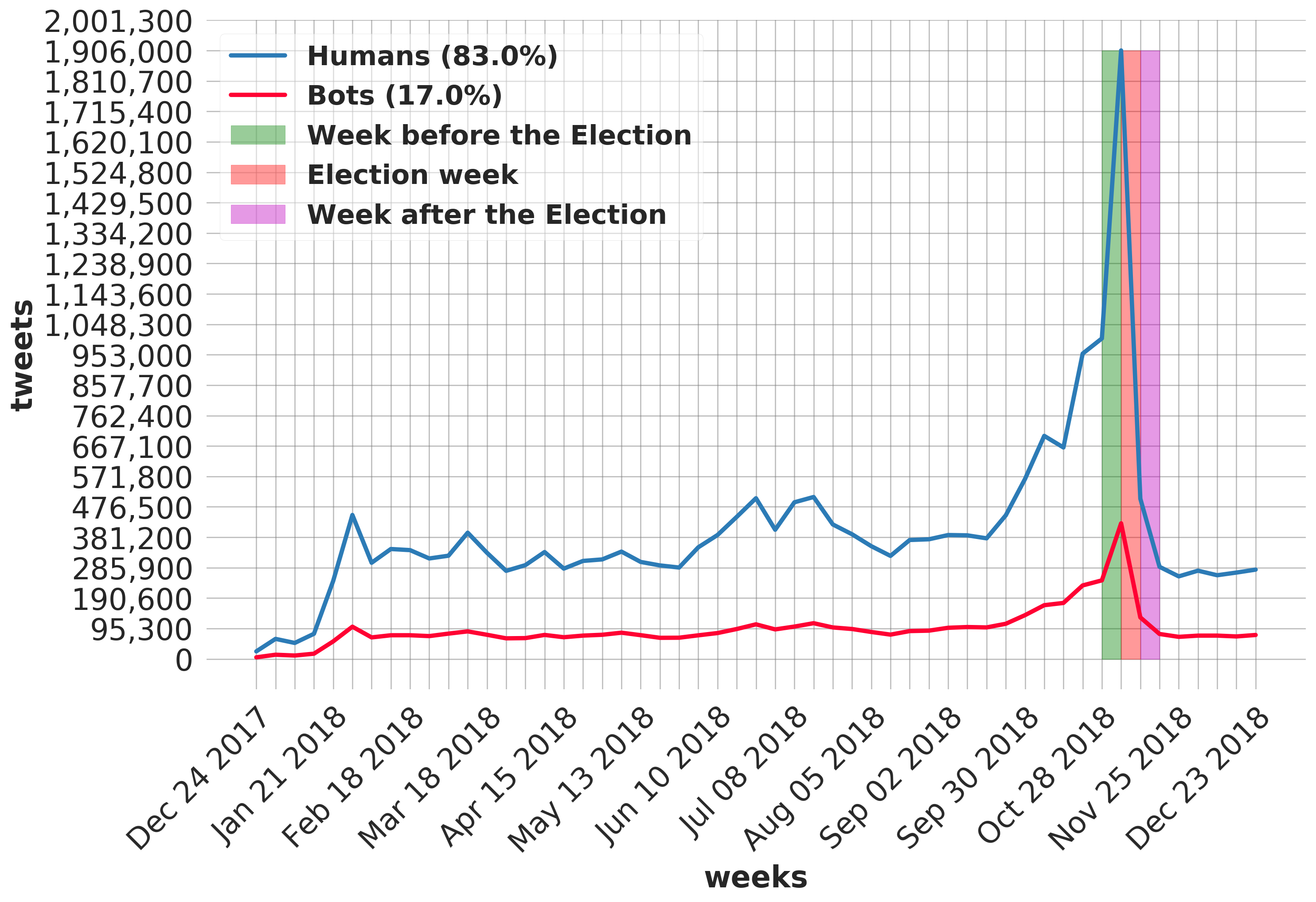}
\caption{Ordinary accounts}
\label{fig:weekly_tweets_less_active}
\end{subfigure}\\
\caption{Tweets shared by hyperactive and ordinary accounts}
\label{fig:most_less_activity}
\end{figure*}

To investigate the behavior of hyperactive accounts, we examine their activity pattern over the whole observation window.
\Cref{fig:weekly_tweets_most_active} portrays their weekly activity.
While in \Cref{fig:weekly_activities_full_tweets} we have noticed that humans generated more tweets than bots in every week, in \Cref{fig:weekly_tweets_most_active}, we recognize a different trend, where hyperactive bots produced more tweets than hyperactive humans over the whole year. 
In \Cref{fig:weekly_tweets_less_active}, we display the number of tweets shared weekly by ordinary accounts.
Differently from \Cref{fig:weekly_tweets_most_active}, the gap between the number of shared tweets by ordinary bots and ordinary humans is remarkable, with the former class of accounts that shared about five times fewer tweets with respect to the latter class. The interested reader can refer to the \textit{Supplement} for the detailed trends and volumes of the disentangled sharing activities (i.e., original tweets, retweets, and replies) for both hyperactive and ordinary users.

\begin{figure*}[h!]
\centering
\begin{subfigure}[t!]{.48\textwidth}
\centering
\includegraphics[width=\columnwidth]{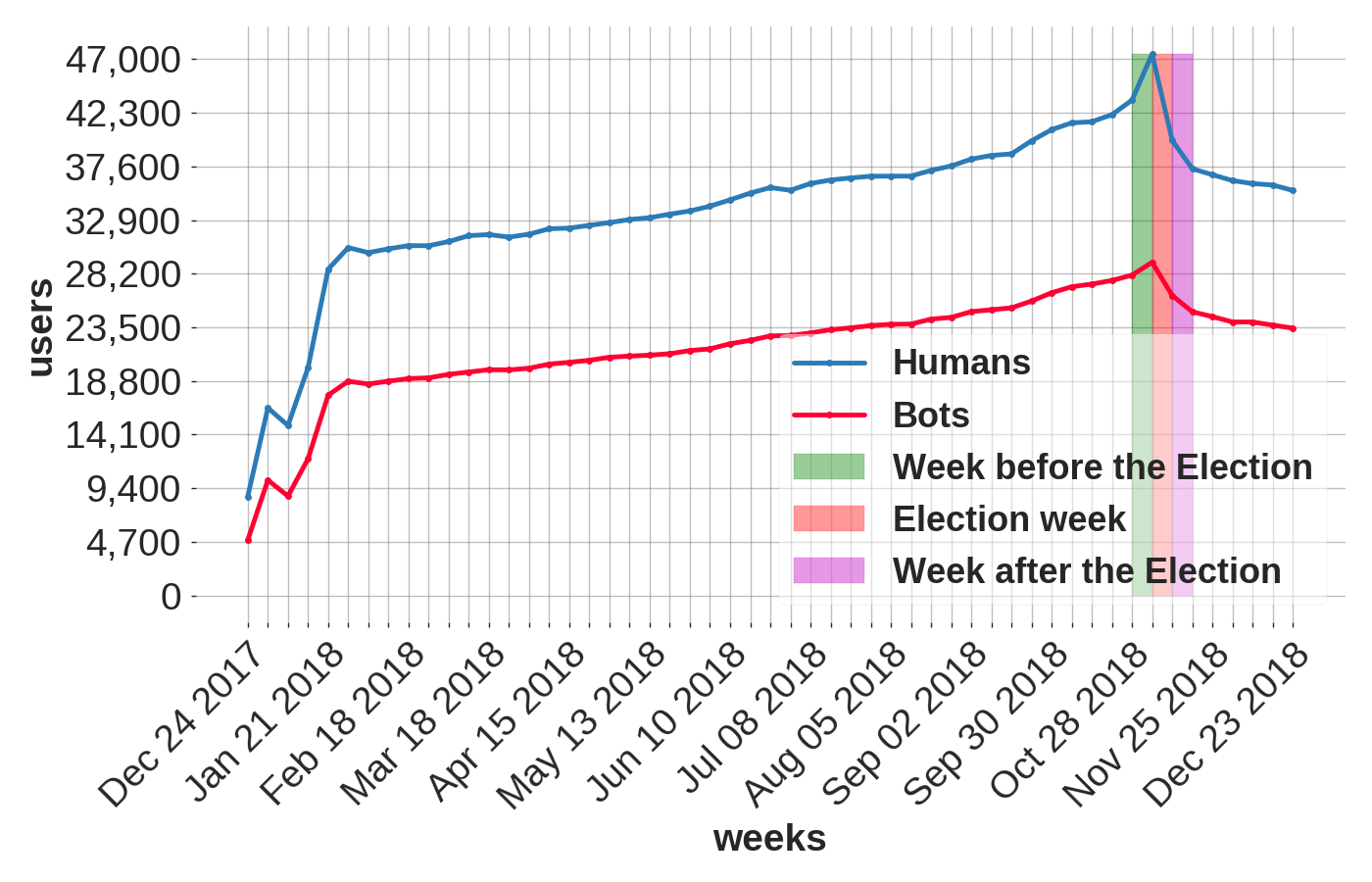}
\caption{Engaged accounts}
\label{fig:weekly_users_most_active}
\end{subfigure}%
\vspace{0.4pt}
\begin{subfigure}[t!]{.48\textwidth}
\centering
\includegraphics[width=\columnwidth]{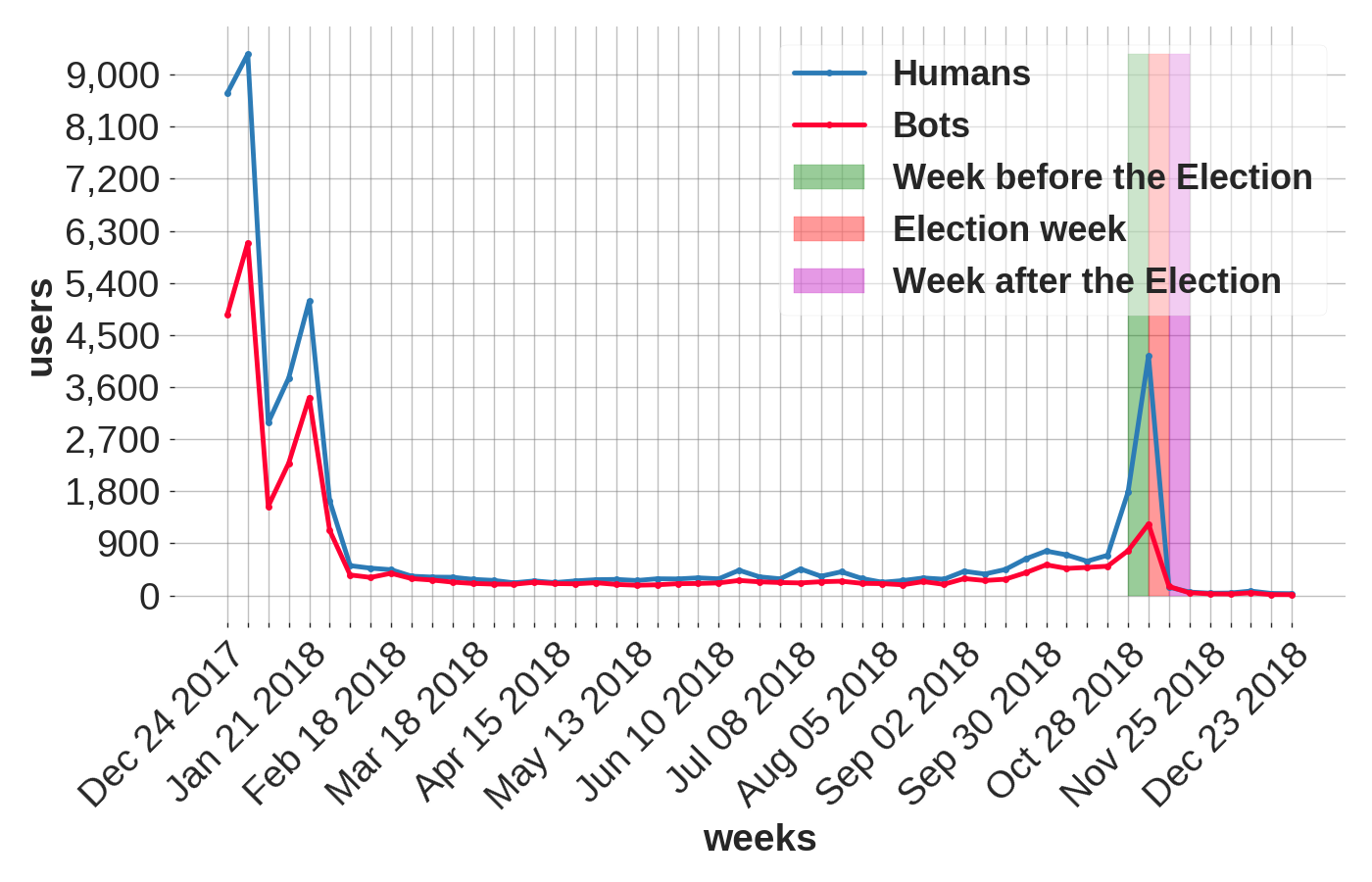}
\caption{New accounts}
\label{fig:weekly_new_users_most_active}
\end{subfigure}\\
\caption{Engaged and new accounts in the Midterm debate within the \textbf{hyperactive} accounts class}
\label{fig:test}
\end{figure*}

Next, we evaluate how the appearance and engagement of the accounts within both classes evolve over time.
In \Cref{fig:weekly_users_most_active}, we show the number of engaged accounts  within the hyperactive class for each week of the year. We recall that an account is considered \textit{engaged} in a certain week if it produced at least one tweet in that week. 
Here, the most noticeable difference with respect to \Cref{fig:weekly_users_full} (where all the accounts are considered) is related to the activity of hyperactive humans.
In fact, as shown in \Cref{fig:weekly_users_most_active}, human activity has a progressive growth over time and a similar pattern to bots activity.
This might suggest that the irregular spikes of activity in \Cref{fig:weekly_users_full}
are mainly caused by ordinary accounts. 
For what pertains to bots, their
weekly activity appears similar to \Cref{fig:weekly_users_full}, which might indicate that bots activity is scheduled similarly for hyperactive and ordinary bot accounts. We further investigate such intuitions in the next paragraphs.

In \Cref{fig:weekly_new_users_most_active}, we show the number of new hyperactive accounts that appear in the Twitter discussion for each week of the year.
 As discussed before, an account is considered \textit{new} when it appears in the Midterm-related conversation for the first time over the observation window. 
 Also in \Cref{fig:weekly_new_users_most_active}, bot and human hyperactive accounts tend to have similar temporal patterns in their appearance within the political discussion.  
 This finding, related to the hyperactive accounts, is in contrast to what we noticed in
\Cref{fig:weekly_new_users_full} (where all the accounts are considered), which further confirms that the different temporal patterns between humans and bots (in both \Cref{fig:weekly_users_full} and \Cref{fig:weekly_new_users_full}) are mainly due to ordinary accounts activity.

Interestingly, and similarly to \Cref{fig:weekly_new_users_full}, in \Cref{fig:weekly_new_users_most_active}, an almost constant number of new hyperactive bot accounts joined the conversation for each week of the observation period, but in the election period, where the number of new hyperactive bot accounts increased significantly.
We also notice that the majority of hyperactive accounts appeared since the end of 2017 (specifically, from December 2017 to February 2018). Such a subset of users (about 5 percent of the accounts) created about 60 percent of the collected tweets. 
This is consistent with the findings reported in the previous paragraph. In particular, here, we recognize that hyperactive accounts were the main responsible for the tweets created by the users that joined the Midterm debate one year before the election. Indeed, of the 79M tweets created by such users, 59M were shared by hyperactive accounts, which, thus, represent the most prolific tweet fabricators. 
Among these accounts, bots played a relevant role by broadcasting almost one-third of the collected tweets. Indeed, 30.6M tweets were created by 19k bot accounts, which means that 2 percent of the accounts in our dataset were bots that shared about 31 percent of the collected tweets. 
This finding becomes even more concerning when considering the pivotal role of the majority of these accounts in the diffusion of narratives related to the QAnon conspiracy theory (14k accounts) and their  involvement in the current U.S. Presidential election debate (16k accounts).



\begin{figure*}[h!]
\centering
\begin{subfigure}[t!]{.48\textwidth}
\centering
\includegraphics[width=\columnwidth]{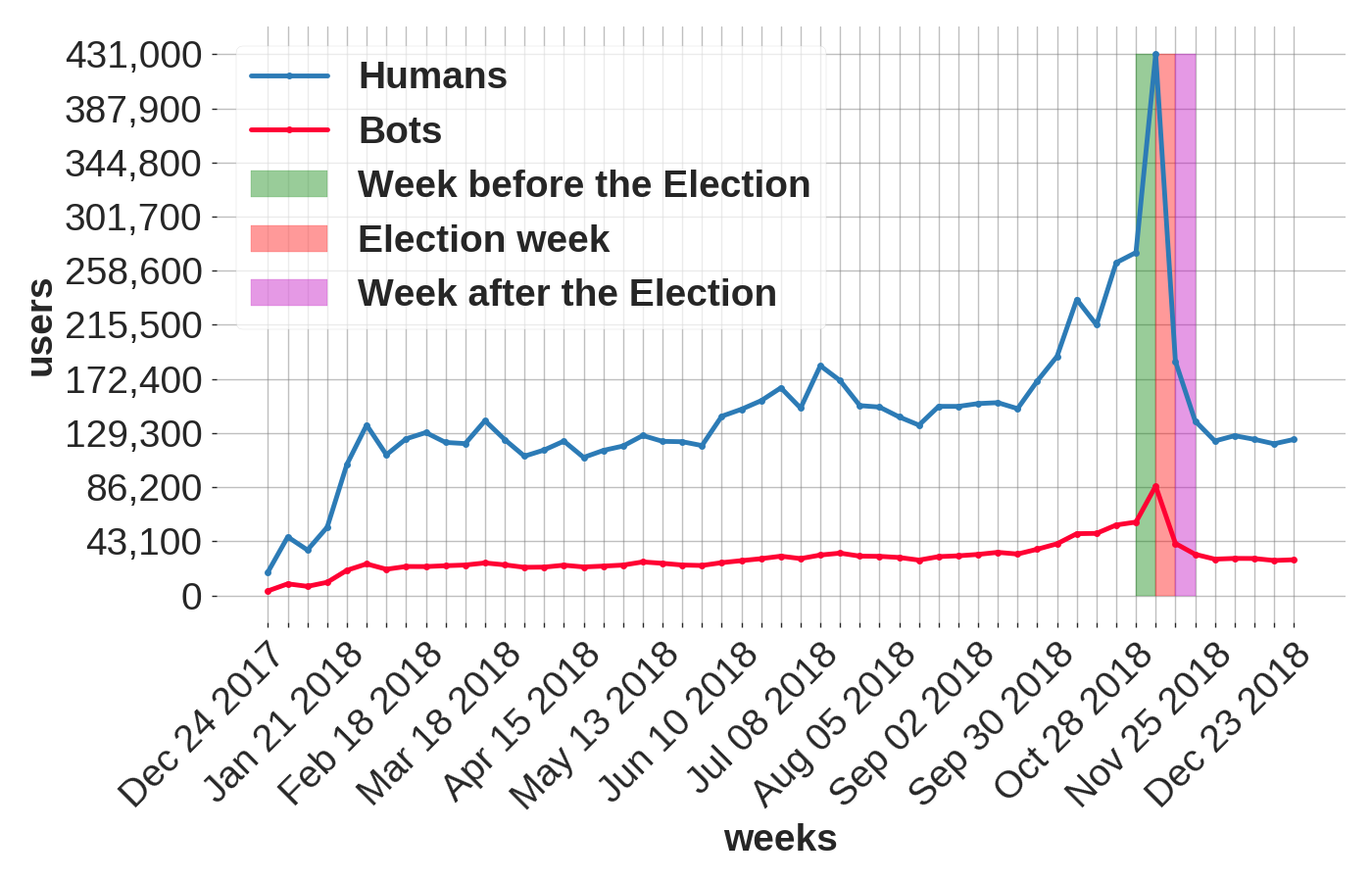}
\caption{Engaged accounts}
\label{fig:weekly_users_less_active}
\end{subfigure}%
\vspace{0.4pt}
\begin{subfigure}[t!]{.48\textwidth}
\centering
\includegraphics[width=\columnwidth]{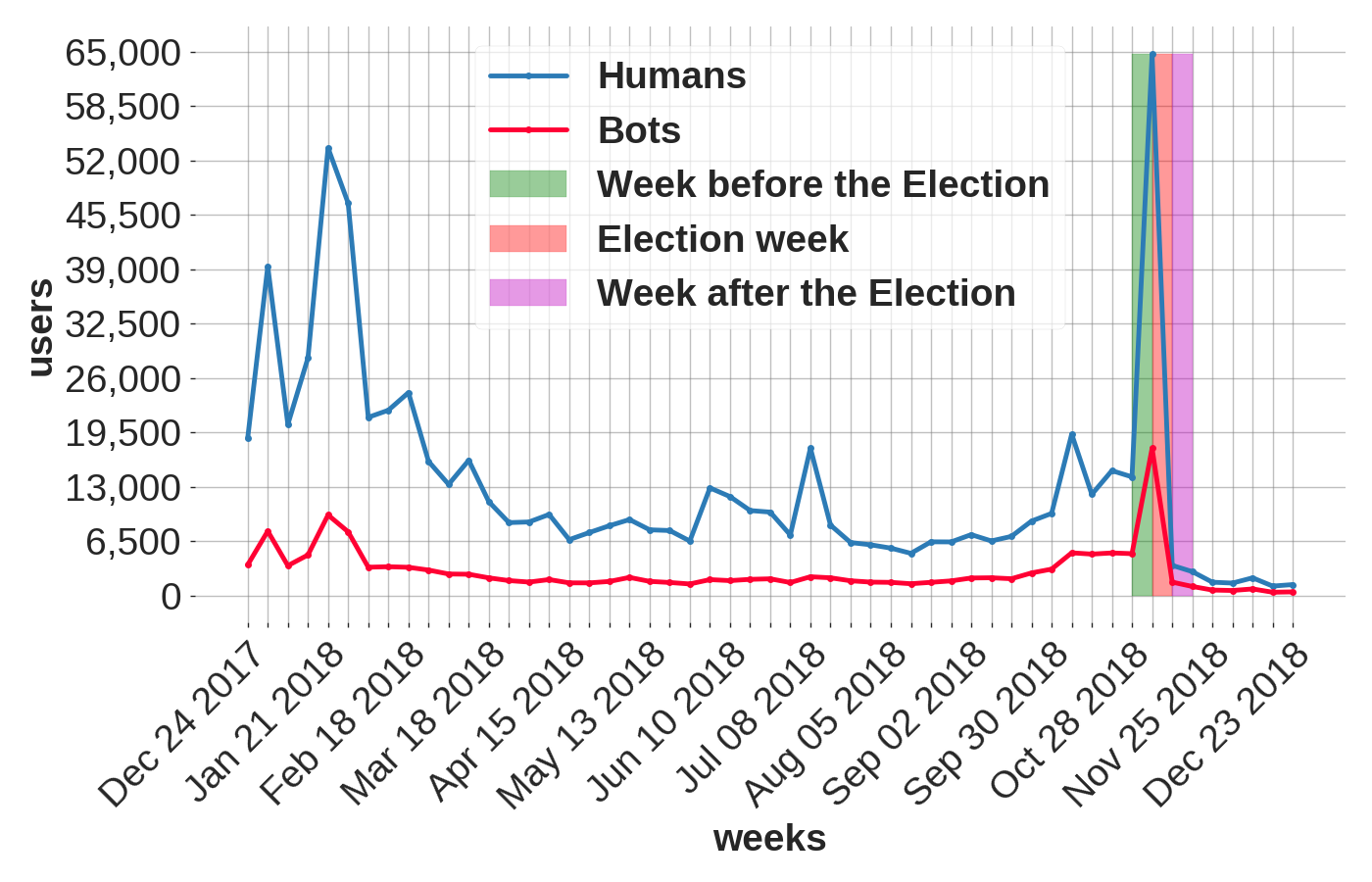}
\caption{New accounts}
\label{fig:weekly_new_users_less_active}
\end{subfigure}\\
\caption{Engaged and new accounts in the Midterm debate within the \textbf{ordinary} accounts class}
\label{fig:less_active_users}
\end{figure*}

Next, we replicate the same evaluation for ordinary accounts. In \Cref{fig:weekly_users_less_active}, we display the number of engaged ordinary accounts for each week of the observation period. Similarly to \Cref{fig:weekly_users_full} (where all the accounts are considered), human accounts present an irregular pattern (with peaks and valleys) of engagement if compared to bots trend, which presents a more controlled and stationary pattern.
The ordinary humans pattern is different from the one observed for hyperactive humans (see \Cref{fig:weekly_users_most_active}), confirming that the non-stationary pattern in \Cref{fig:weekly_users_full} is mainly caused by ordinary human accounts.

In \Cref{fig:weekly_new_users_less_active}, we depict the number of new ordinary accounts that appeared in the Twitter discussion for each week of the year. 
Similarly to \Cref{fig:weekly_new_users_full} and \Cref{fig:weekly_new_users_most_active}, bots entered into the debate at a constant and lower rate with respect to human accounts.
Also, the majority of ordinary accounts joined the discussion at the beginning of 2018 and during the election month. Interestingly, by disentangling the shared tweets, we notice different activity patterns among the bots injected from December 2017 to the first two weeks of February 2018 (\textit{old} bots) and those appearing the month around the election day (\textit{election} bots).
In \Cref{table:old_vs_election_less_active_bots}, we detail how these two classes of accounts distributed their sharing activities. The main difference is related to the creation of original content and the usage of retweets, similarly to \Cref{table:old_vs_election_bots} (where all the accounts are considered). Here, the discrepancy between old and election bots is even more pronounced, with election bots that performed re-sharing activities only 58 percent of the time, while sharing original tweets 37 percent of the time. This result further confirms the different approaches used by old and election bots and highlights their diverse operational tasks.

\begin{table}[h]
    \caption{Sharing activities of ordinary bot accounts}
    \centering
    \begin{tabular}{l|l|ll}
    
    {} &      \textbf{Old bots} &    \textbf{Election bots}\\
    \hline
    \textbf{Tweets       } &      
3,553,531 &   37,713\\
    \hline
    \textbf{Original tweets          } &  16\% &  37\%\\
    \hline
    \textbf{Retweets          }  &  80\% &  58\%\\
    \hline
    \textbf{Replies          }&  4\% &  5\%\\
    \end{tabular}
    \label{table:old_vs_election_less_active_bots}
\end{table}

\begin{figure*}[h!]
\centering
\begin{subfigure}[t!]{.48\textwidth}
\centering
\includegraphics[width=\columnwidth]{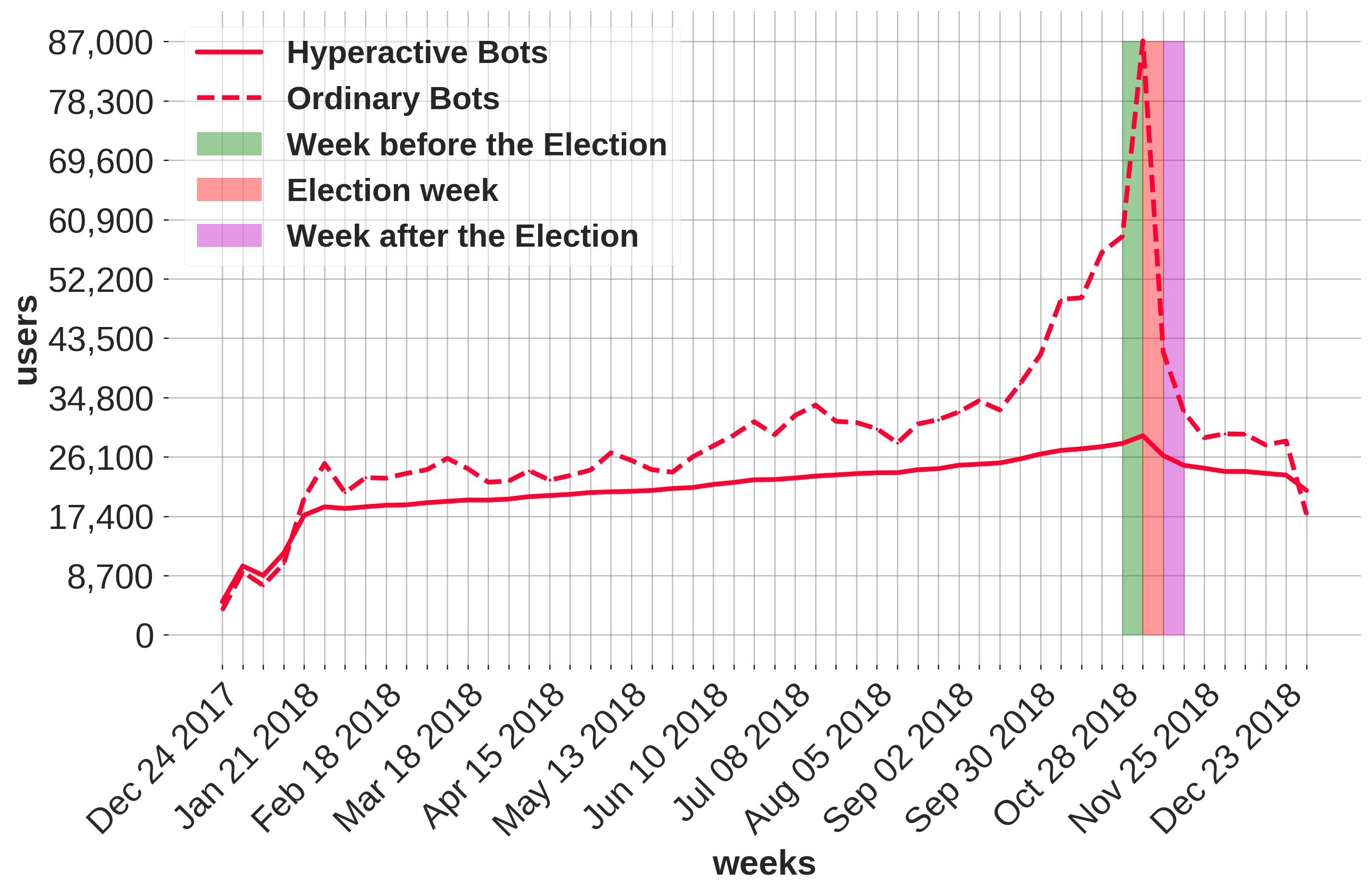}
\caption{Engaged bots}
\label{fig:weekly_bots_most_less_active}
\end{subfigure}%
\vspace{0.4pt}
\begin{subfigure}[t!]{.48\textwidth}
\centering
\includegraphics[width=\columnwidth]{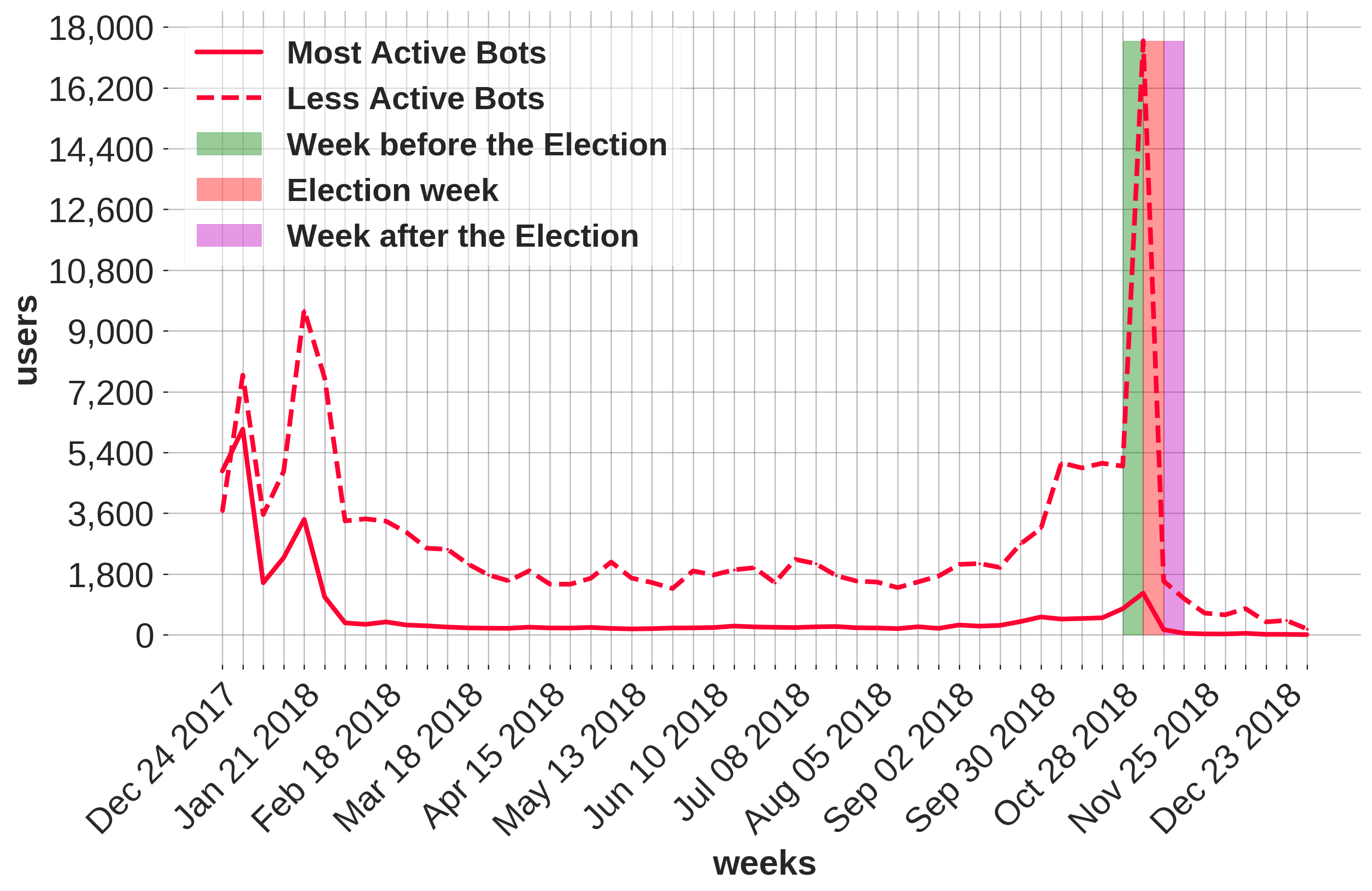}
\caption{New bots}
\label{fig:weekly_new_bots_most_less_active}
\end{subfigure}\\
\caption{Engaged and new bots within Midterm debate}
\label{fig:weekly_ma_la_bots}
\end{figure*}

To further examine the different behavior between hyperactive and ordinary bots, in \Cref{fig:weekly_ma_la_bots}, we compare the activity of such classes of accounts. In particular, \Cref{fig:weekly_bots_most_less_active} depicts the number of engaged bots for each week of the year. It can be noticed how the engagement of hyperactive bot accounts presents a more steady pattern, with an increasing number of accounts progressively engaged as the election approached, if compared to the trend of ordinary bots, which presents a less stationary pattern of engagement and a more pronounced growth in the month prior to the election. 

This is also reflected in \Cref{fig:weekly_new_bots_most_less_active}, which displays the number of new bots accounts that joined the Midterm discussion over the year. From \Cref{fig:weekly_new_bots_most_less_active}, we can observe how a significant number of bots were active in the conversation since the beginning of the year and new bots were weekly injected over the months preceding the election. It can be noticed how a small and constant number of hyperactive bots joined the conversation weekly, while a larger set of ordinary bots was injected over the year and especially during the election week.

\paragraph{RQ4: Interactions and embeddedness of bot and human accounts}

In this paragraph, we explore the interactions between human and bot accounts. In particular, we examine the interplay in terms of retweets (re-sharing of other users' content) and replies (respond to other users' content) exchanged between these two classes of accounts over the whole period of observation. We also investigate the embeddedness of both bots and humans within the social network with the objective of measuring their centrality in the online discussion.

\begin{figure}[h!]
\vspace{+.5cm}
    \centering
    \includegraphics[width=0.6\linewidth]{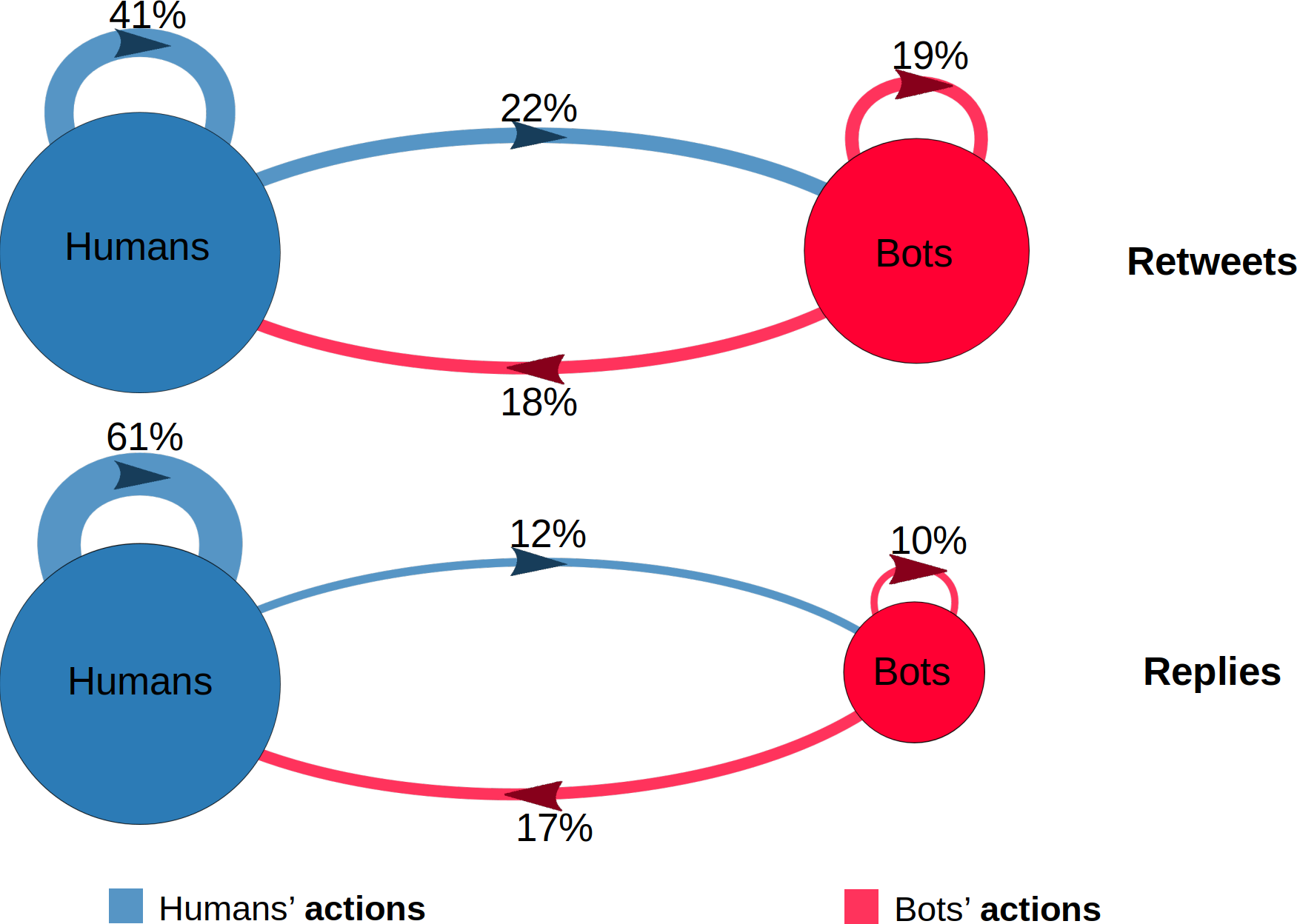}
    \vspace{+.5cm}
    \caption{Retweet and reply interactions between humans and bots}
    \label{fig:rt_rp_nets_overall}
\end{figure}

\Cref{fig:rt_rp_nets_overall} shows the interactions in terms of retweets and replies between the two classes of accounts, where each edge reports the volume of interactions (in terms of percentage) between each
group. 
The \textit{source} of the edge represents the user that re-shared (i.e., retweeted) the tweet of a \textit{target} user.
Node size is proportional to the percentage of accounts in each group for the considered sharing activity,
while edge size is proportional to the percentage of interactions
between each group.
On the one hand, we observe that human content is more re-shared (59 percent of the retweets) than bot-generated content (41 percent of the retweets). Interestingly, bots equally distribute their retweets towards human- and bot-generated content, which might indicate a strategy operated by bots to engage humans and possibly gain their endorsement \cite{luceri2019red}.
It is important to note that more than one over three human retweets is a re-share of content produced by bots. Such indiscriminate re-sharing, which allegedly occurs because of humans' ineptitude in recognizing inaccurate sources of information, represents a complex issue that might have dramatic repercussions in the spread of misinformation \cite{bessi2016social,shao2018spread,vosoughi2018spread}.

On the other hand, we recognize that humans performed the majority of replies and only a small percentage of those (around 12 percent) were in response to bot-generated content, which is aligned with the findings of Bessi and Ferrara \cite{bessi2016social} related to the 2016 U.S. Presidential election. Bots, in turn, focused their replies on the interaction with humans, which appears in contrast with the 2016 pattern \cite{bessi2016social}, where bots interacted with other bot accounts more than with human users. We hypothesize that bots have become increasingly advanced to initiate conversations with humans \cite{luceri2019evolution}, but allegedly not sufficiently sophisticated to convincingly involve humans in their discussion. Indeed, bots received much less interaction (in terms of reply) from humans with respect to the retweet interaction. 

Next, we explore the embeddedness of both bots and humans within the social network by also considering (i) the extent of their sharing activity (hyperactive vs. ordinary accounts) and (ii) the time of their appearance within the Midterm debate. We recall that we indicate an account as \textit{old} if it shared the first Midterm-related tweet by the first two weeks of February 2018. Otherwise, we here denote the account as \textit{recent}. As detailed before, this choice is motivated by the evidence that a significant number of accounts joined the Twitter conversation from the early stage of the Midterm debate (see Figure 5b) and, in turn, generated a disproportionate volume of tweets if compared to the recent ones.

\begin{figure*}[h!]
    \centering
        \begin{subfigure}[t!]{.495\textwidth}
        \centering
            \includegraphics[width=\textwidth]{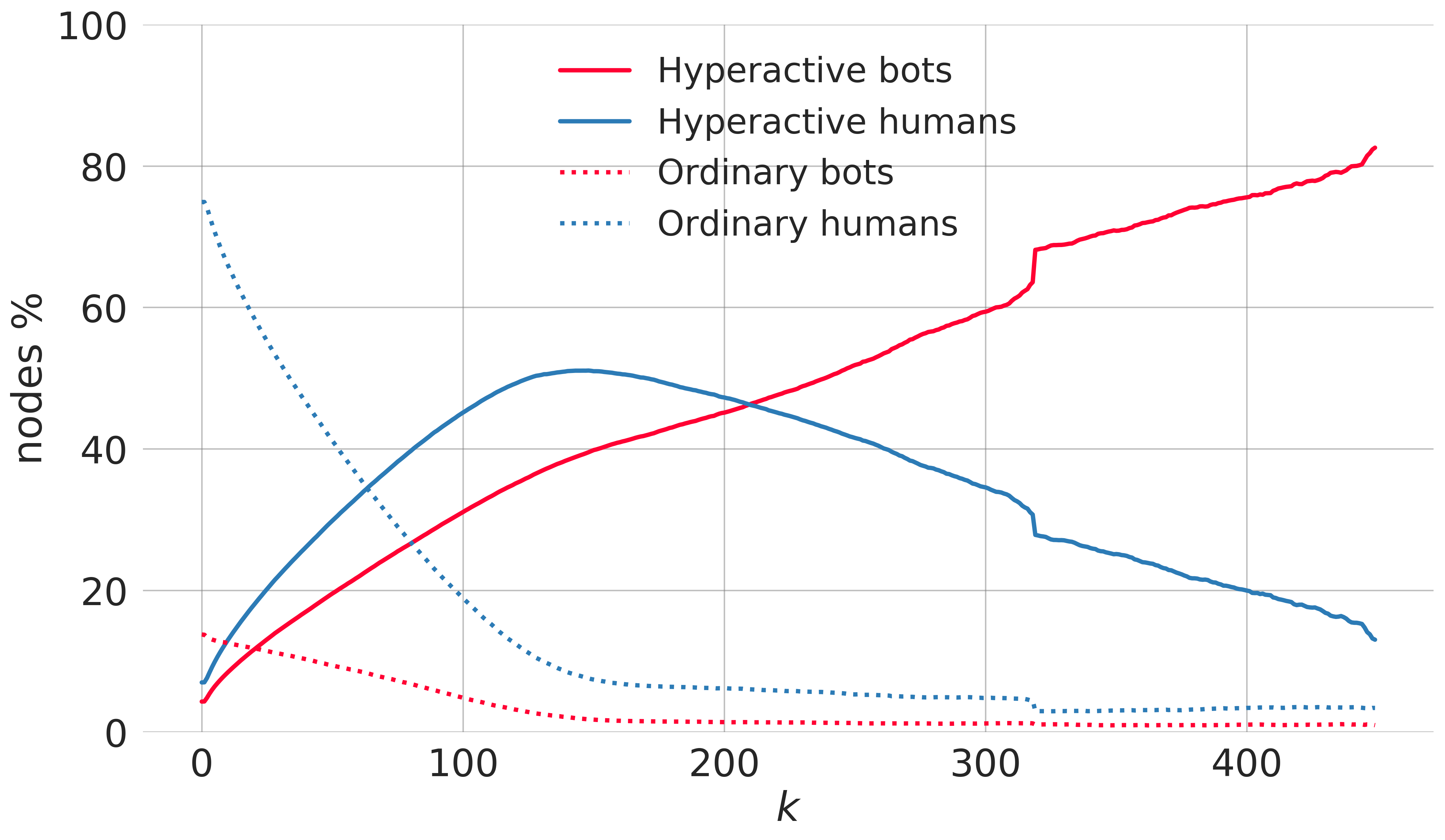}
            \caption{Hyperactive vs. ordinary accounts}
            \label{fig:k_core_rt_mixed}
        \end{subfigure}%
    \vspace{0.4pt}
        \begin{subfigure}[t!]{.495\textwidth}
        \centering
            \includegraphics[width=\textwidth]{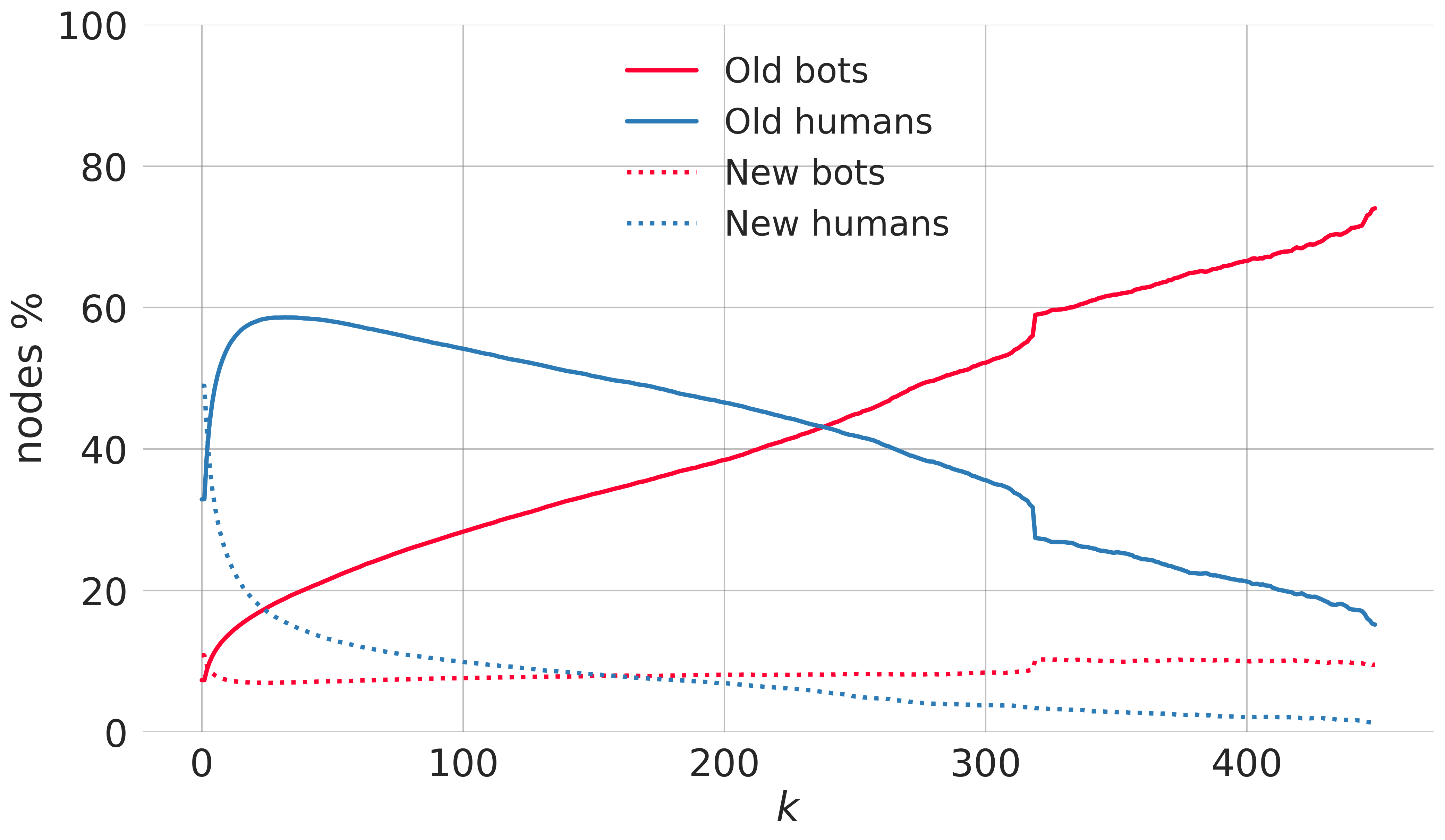}
            \caption{Old vs. recent accounts}
            \label{fig:k_core_old_new}
        \end{subfigure}\\
    \caption{$k$-core decomposition of the retweet network for bot and human accounts}
    \label{fig:k_core_rt_active_old}
\end{figure*}

To perform such analysis, in \Cref{fig:k_core_rt_active_old}, we depict the $k$-core decomposition of the retweet network for the two classes of accounts (bots and humans). 
A retweet network is a directed weighted graph that represents the retweet interactions between users, where nodes represent users and edges represent retweets among them.
We extracted the $k$-cores from the retweet network by varying $k$ in the range between 0 and 500 (no variations can be observed after this value). \Cref{fig:k_core_rt_active_old} displays the percentage of hyperactive vs. ordinary (resp. old vs. recent) users as a function of $k$, i.e., for every $k$-core we compute the proportion of accounts of each group within the accounts in the $k$-core.

From \Cref{fig:k_core_rt_mixed}, we can notice that the fraction of hyperactive bots increases with $k$, while the percentages related to ordinary accounts drop with $k$. It appears evident how hyperactive users, and in particular bot accounts, populate the most central position within the retweet network. To the contrary, within the ordinary accounts set, human users appear to hold a more central position within the retweet network with respect to bot accounts. 
\Cref{fig:k_core_old_new} depicts the $k$-core decomposition of recent and old accounts. It can be noticed that old bots represent the group of accounts most embedded within the retweet network, which indicates that they are more deeply connected with respect to old humans and recent accounts. Interestingly, we observe that, as $k$ grows, the fraction of humans drops, whereas the percentage of recent bots remains almost steady. This suggests that a fraction of recent bots populate the most central area of the social network, which, in turn, shows their capability of infiltrating the online discussion and rapidly gaining trust and reputation even in a relatively short time window.




\section*{Discussion and Conclusion}
Bots represent one of the most recognized threats to the integrity of social media. The understanding of how bot accounts operate in social platforms and infiltrate into crucial discussions is of pivotal importance to enable prompt and precautionary detection of coordinated campaigns.
Further,
it is also fundamental measuring and controlling the extent of bots interactions with organic users to assess the impact of their manipulation attempts.
Along these research directions, in this paper, we examined how bots strategically performed their online activity over time during the year approaching the 2018 U.S. Midterm elections, as well as the effects of their interplay with human users. 

We observed how bots reproduced humans' activity patterns for every sharing activity (i.e., original tweet, retweet, and reply) since the year before the election and we recognized their propensity of flooding online platforms with retweets. However, an increasing number of bots created with the purpose of sharing original content was strategically injected into the online conversation as the election day approached. 
Interestingly, we discovered that a significant fraction of bots (about one-third of the identified automated accounts) started pushing Midterm-related content even one year before the election, generating a conspicuous volume of tweets (34.2 million) over the whole observation period. Alarmingly, this 
group of prolific tweet fabricators encompasses a small set of hyperactive bots (19k accounts) that acted continuously over the whole 2018 broadcasting 30.6M tweets and most of them (16k accounts) are also operating in the 2020 U.S. Presidential debate at the time of this writing. This finding becomes even more concerning when considering that hyperactive accounts promoted the vast majority of conspiratorial narratives related to the QAnon movement. 
 
Further, we noticed that bots entered into the Midterm debate at a more regular and lower rate with respect to human accounts. Specifically, a constant number of bot accounts were injected and participated in the Midterm conversation every week of the year, suggesting the implementation of a scheduled and cautious strategy to infiltrate bots within the online discussion while avoiding their detection. Besides these bots, a relevant fraction of automated accounts have been created a few months prior to the election (from July to November 2018) and participated in the related discussion, which might indicate that such accounts have been purposely injected for operating into the Midterm elections debate. Interestingly, these bots generated significantly more original content, while decreasing the usage of the retweet, with respect to \textit{older} bot accounts.
This was particularly evident for bot accounts with a moderate engagement in the Midterms debate (i.e., ordinary accounts), which suggests that diverse operational tasks were deployed on more recent and less recent bot accounts, also based on their degree of engagement within the conversation (i.e., hyperactive vs. ordinary accounts).

Finally, bots resulted to be more deeply connected in the social network with respect to humans, especially the hyperactive and less recent ones. It stands out how a fraction of more recent bot accounts have been capable of infiltrating the online discussion to the extent to rapidly gain trust and stand in a central position of the social network in a relatively short time window.
In terms of interactions, bots equally distributed their retweets towards human- and bot-generated content, while their replies were more focused on the interplay with humans. For what pertains to human accounts, it is alarming noting that one over three human’s retweets was a re-share of content produced by bots, which further highlights the need for intervention to avoid the diffusion of misinformation. However, our results also reveal that bots received less interaction from humans in terms of replies (if compared to retweets), which might represent an encouraging finding when considering users’ awareness of and response to the activity of malicious entities on social media. 
 
 
Concluding, our study confirms how the detection of manipulation campaigns is a complex and challenging task, even when suspected malicious accounts are monitored over an extended observation window. Indeed, we have shown that a pivotal set of bots mimicked human sharing activities by emulating the volume and temporal patterns of their narratives several months before the voting event while dominating the debate and infiltrating the most influential areas of the social network. 
These results should be a cause of concern when considering the persuasive power of social media, especially in the political context, where the upcoming 2020 U.S. Presidential Election represents a global testing ground for our democracy.

However, our analysis revealed novel patterns (e.g., the increasing injection of bots aimed at creating original content as the election approached) and actionable insights (e.g., hyperactive bot accounts started their sharing activity one year before the election and were responsible for fueling conspiratorial narratives) that can empower the identification of malicious entities and, accordingly, contribute to the
detection of coordinated campaigns.
These findings, and corresponding analysis, will be validated in our future work by encompassing both other voting events discussions and social conversations on diverse contexts (e.g., public health). 

\section*{\large Acknowledgments}
The authors are funded by the European Institute of Innovation \& Technology via the project Virtual Machina, and by the Swiss National Science Foundation via the project Late teenagers Online Information Search (LOIS). 
The authors are grateful to Emilio Ferrara's team at USC for releasing the Twitter 2020 U.S. Presidential election dataset, and to Goran Muric (USC) for his support with the dataset.

\bibliography{biblio}

\bibliographystyle{splncs04} 

\section*{Supplement}

\paragraph{Accounts involved in both 2018 and 2020 debates
}
In this paragraph, we provide detailed information about the accounts that are analyzed in this manuscript and are also involved, at the time of this writing (mid-October 2020), in the 2020 U.S. Presidential election debate on Twitter.
We refer to this subset of accounts as \textbf{users 2018 \& 2020}, while we denote the whole set of accounts analyzed in this paper as \textbf{users 2018}. 

To attain the set of users engaged in the 2020 U.S. Presidential election debate, we leverage the Twitter dataset collected by \cite{chen2020election2020}. Among the set of users involved in the more recent debate, we spot only those that are analyzed in our paper and, accordingly, identify 582,349 \textbf{users 2018 \& 2020}. In Table \ref{table:users_2018_2020}, we detail the statistics of such users and compare them with the \textbf{users 2018}, whereas Table \ref{table:old_users_2018_2020} shows further details related to \textit{old} \textbf{users 2018 \& 2020}.

\begin{table}[h!]
\vspace{0.5cm}
\footnotesize
\centering
\caption{Users 2018 \& 2020 and users 2018 statistics}
\vspace{.3cm}
\begin{tabular}{l|l|ll|ll}                                      & \textbf{All} & \textbf{Bots} & \textbf{Humans} & \textbf{Hyper.} & \textbf{Ordinary}  \\ 
\hline
\textbf{Users 2018 \& 2020}           & 582k         & 93k           & 489k            & 60k                  & 467k                \\
(Users 2018)                 & (943k)       & (184k)        & (759k)          & (84k)                & (698k)                \\ 
\hline
\textbf{Tweets by users 2018 \& 2020} & 77M          & 33M           & 44M             & 56M                  & 21M                      \\
(Tweets by users 2018)     & (98M)        & (42M)         & (56M)           & (71M)                & (27M)                    
\end{tabular}
\label{table:users_2018_2020}
\end{table}

\begin{table}[h!]
\vspace{0.3cm}
\centering
\caption{Old users 2018 \& 2020 statistics}
\vspace{0.3cm}
\begin{tabular}{l|l|l}
\multicolumn{1}{l}{}           & \multicolumn{2}{c}{\textbf{Old}}          \\
                               & \textbf{Hyperactive} & \textbf{Ordinary}  \\ 
\hline
\textbf{Accounts 2018 \& 2020} & 42k                  & 207k               \\
(Accounts 2018)                & (51k)                & (266k)             \\ 
\hline
\textbf{Bots 2018 \& 2020}     & 16k                  & 28k                \\
(Bots 2018)                    & (19k)                & (40k)              \\
\textbf{Humans 2018 \& 2020}   & 26k                  & 179k               \\
(Humans 2018)                  & (32k)                & (226k)            
\end{tabular}
\label{table:old_users_2018_2020}
\vspace{.5cm}
\end{table}

\paragraph{Suspended Accounts}
In this paragraph, we examine the correlation between account suspension and bot scores.
Following the methodology proposed in \cite{ferrara2020types}, we leverage the annotations of verified and suspended accounts and compare their bot scores.

\begin{figure}[h!]
\vspace{.5cm}
    \centering
    \includegraphics[width=0.7\linewidth]{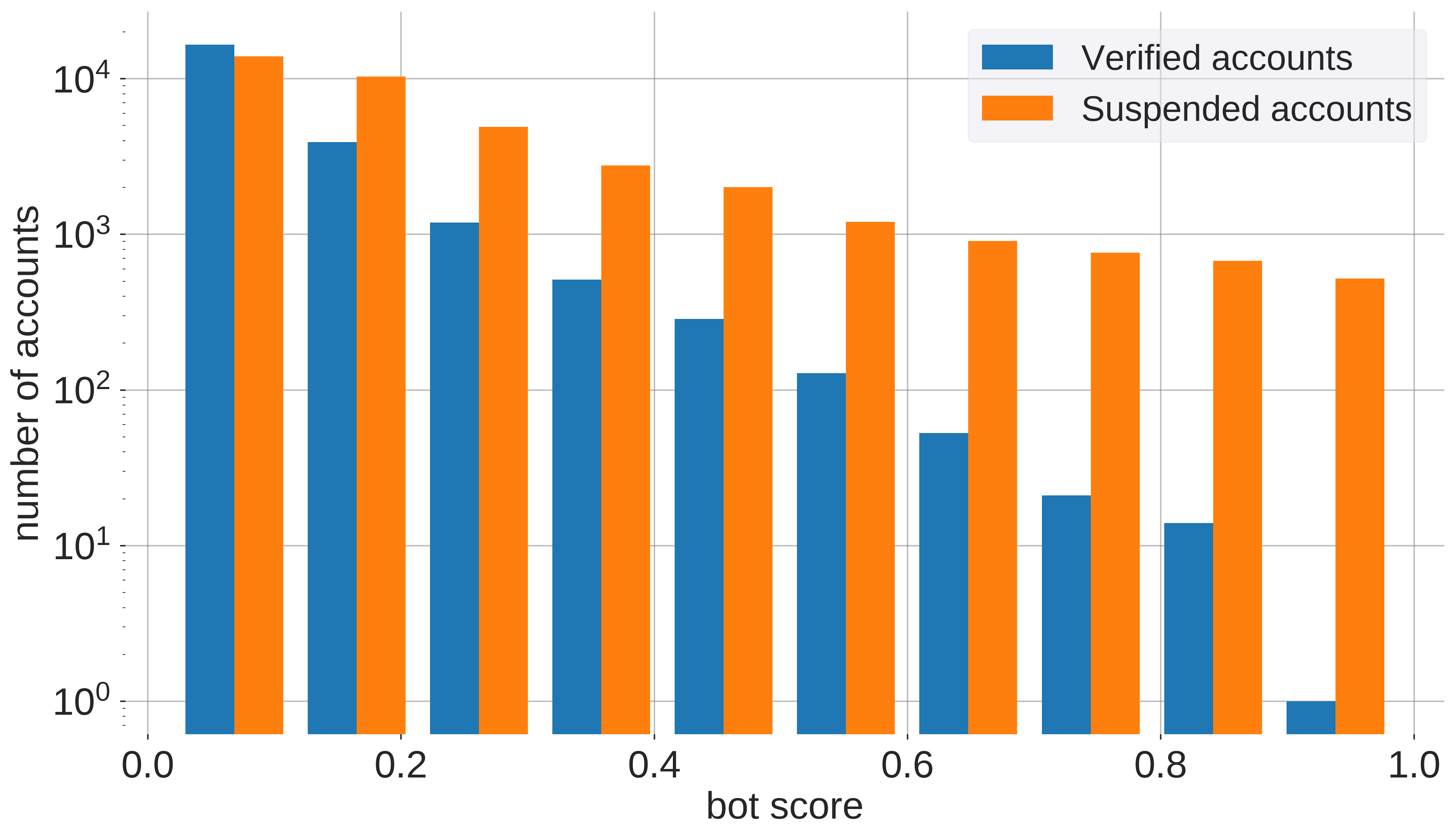}
    \caption{Distribution of bot scores for verified and suspended accounts}
    \label{fig:susp_verif_scores}
\end{figure}

In Figure \ref{fig:susp_verif_scores}, we depict the bot scores distributions of the 38k suspended accounts and 22.7k verified accounts within our dataset.
We observe two different distributions, as statistically confirmed by the Mann-Whitney rank test (\textit{p}-value $<$0.001).
Indeed, bot scores of the suspended accounts are more widely distributed with respect to those of the verified accounts, which have about 80 percent of the users with a bot score lower than 0.15. To the contrary, more than half of the suspended accounts present bots scores higher than 0.15.
It should be noticed that account suspension from Twitter may happen for several reasons, not only related to the degree of automation of the account. Therefore, it was expected to also have suspended accounts with low bot scores. However, it can be noted how thousands of suspended accounts achieve a high level of automation (i.e., high bot scores), while the number of verified accounts drops as the bot score increases. This, besides confirming a correlation between account suspension and the likelihood of automation, supports the accounts classification performed via Botometer \cite{ferrara2020types}.

\paragraph{Hyperactive accounts}
\Cref{table:active_filtered_properties} summarizes the properties related to both hyperactive bot and human accounts and the volume of their shared tweets. 

\begin{table*}[h!]
\small
    \setlength{\tabcolsep}{10pt}
    \vspace{.25cm}
    \caption{Hyperactive accounts properties}
    \vspace{.1cm}
    \centering
    \begin{tabular}{l|cc}
               \textbf{} &       \textbf{Value} & \textbf{\% (\% within hyperactive accounts)} \\
               \hline
         \textbf{Total Accounts} &      83,701 &             8.87\% (100\%) \\
           \textbf{Bot Accounts} &      31,772 &             3.37\% (37.96\%) \\
        \textbf{Humans Accounts} &      51,929 &             5.51\% (62.04\%) \\
                \hline
           \textbf{Total Tweets} &  70,737,005 &            72.09\% (100\%) \\
             \textbf{Bot Tweets} &  37,169,582 &            37.88\% (52.55\%) \\
           \textbf{Human Tweets} &  33,567,423 &            34.21\% (47.45\%) \\
    \end{tabular}
        \vspace{.25cm}
    \label{table:active_filtered_properties}
\end{table*}

\begin{figure}[h!]
    \vspace{.3cm}
    \makebox[\textwidth][c]{
    \centering
        \includegraphics[width=\linewidth]{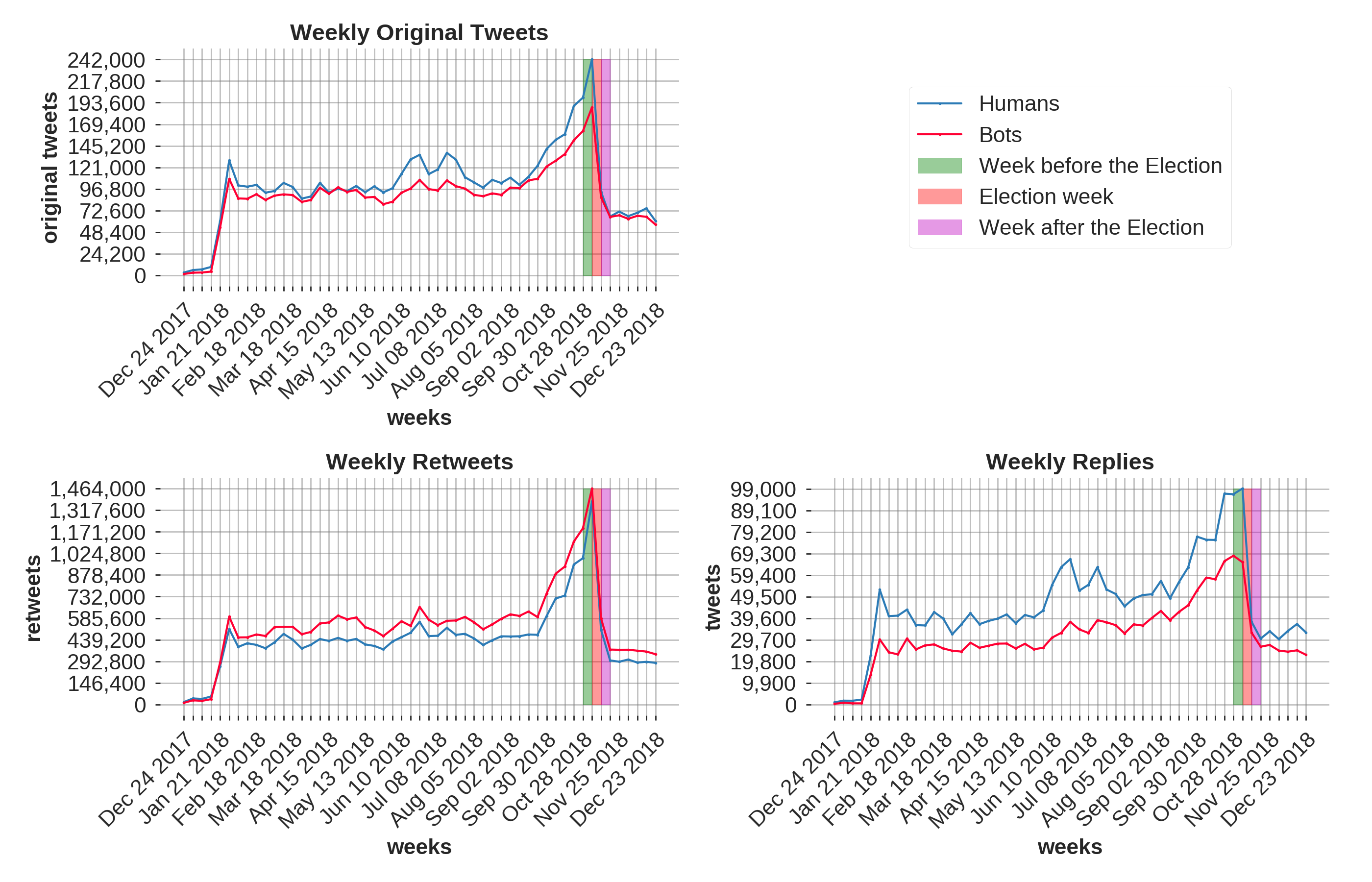}
    }
    \caption{Weekly sharing activities for Hyperactive Accounts}
    \label{fig:weekly_multi_users_most_active}
\end{figure}

\Cref{fig:weekly_multi_users_most_active} displays the sharing activities of hyperactive accounts over the observation period.   

\paragraph{Ordinary accounts}

\Cref{table:less_active_filtered_properties} summarizes the properties related to both ordinary bot and human accounts and the volume of their shared tweets.   

\begin{table}[h!]
\small
    \vspace{.3cm}
    \setlength{\tabcolsep}{10pt}
    \caption{Ordinary accounts properties}
        \vspace{.3cm}
    \centering
    \begin{tabular}{l|cc}
               \textbf{} &       \textbf{Value} & \textbf{\% (\% within ordinary accounts)} \\
            \hline
            \textbf{Total Accounts}   &   698,427    &       74.05\% (100.0\%) \\
            \textbf{Bot Accounts}   &     118,503    &       12.56\% (16.97\%) \\
            \textbf{Human Accounts} &     579,924    &       61.49\% (83.03\%) \\
            \hline
            \textbf{Total Tweets}     &  27,225,609 &  27.75\% (100.0\%) \\
            \textbf{Bot Tweets}       &   5,182,417 &   5.28\% (19.04\%) \\
            \textbf{Human Tweets}     &  22,043,192 &  22.46\% (80.96\%) \\
    \end{tabular}
    \label{table:less_active_filtered_properties}
    \vspace{.3cm}
\end{table}

\begin{figure}[h!]
    \vspace{.3cm}
   \centering
   \includegraphics[width=\linewidth]{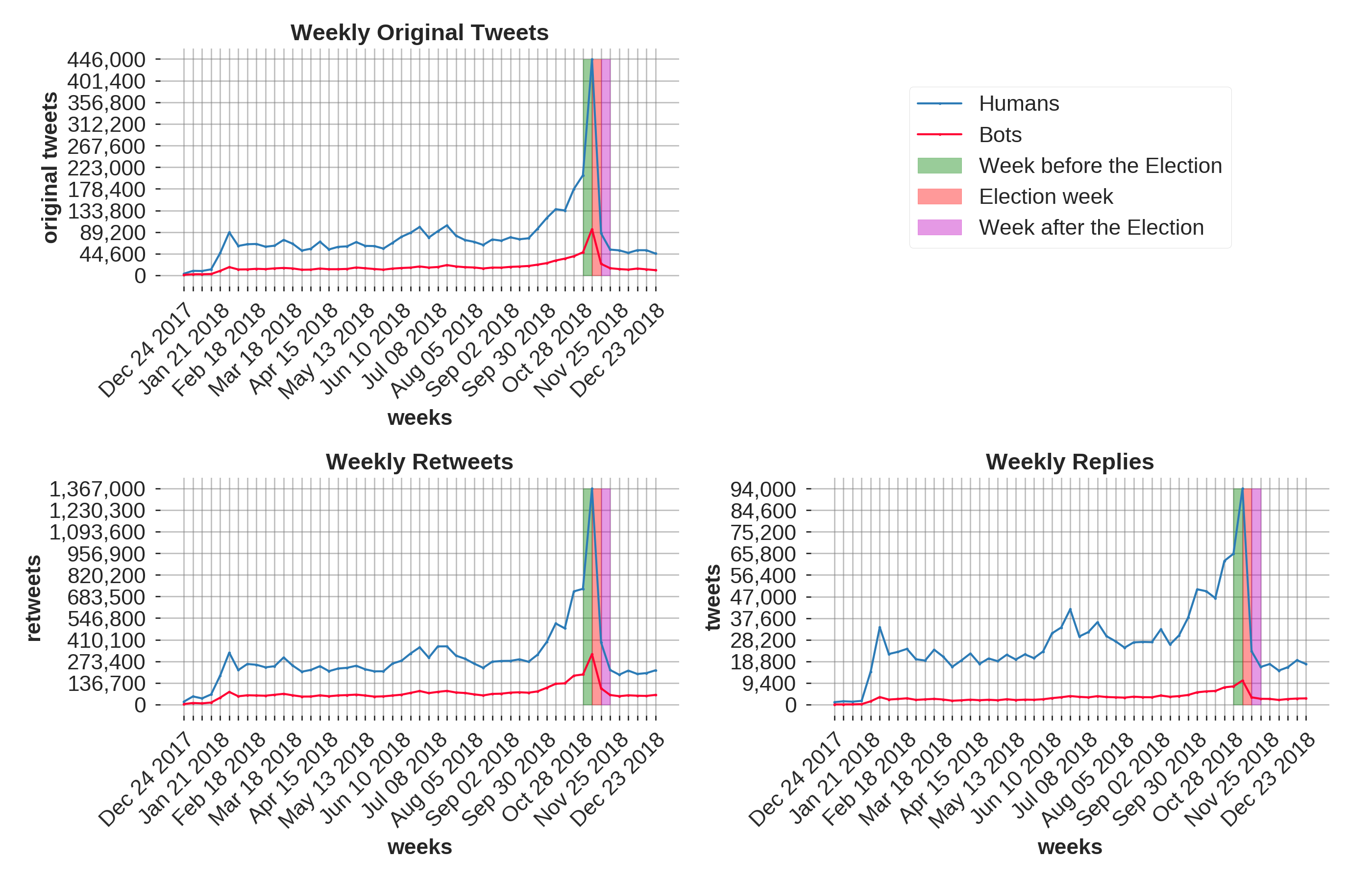}
    \caption{Weekly sharing activities for ordinary accounts}
    \label{fig:weekly_multi_users_less_active}
\end{figure}

\Cref{fig:weekly_multi_users_less_active} displays the sharing activities of ordinary accounts over the observation period.  

\paragraph{One-tweet accounts}
\Cref{table:one_tweet_filtered_properties} summarizes the properties related to the accounts that shared only one tweet.   

\begin{table}[h!]
\small
\vspace{.25cm}
    \setlength{\tabcolsep}{10pt}
    \vspace{.1cm}
    \caption{Properties of the accounts that shared only one tweet}
    \centering
    \begin{tabular}{l|cc}
               \textbf{} &       \textbf{Value} & \textbf{\% (\% within one-tweet accounts)} \\
            \hline
        Total Accounts   &  160,998 &  17.07\% (100.0\%) \\
          Bot Accounts   &   34,270 &   3.63\% (21.29\%) \\
        Human Accounts &  126,728 &  13.44\% (78.71\%) \\
            \hline
        Total Tweets &  160,998 &   0.16\% (100.0\%) \\
          Bot Tweets &   34,270 &   0.03\% (21.29\%) \\
        Human Tweets &  126,728 &   0.13\% (78.71\%) \\
    \end{tabular}
        \vspace{.25cm}
    \label{table:one_tweet_filtered_properties}
\end{table}

\paragraph{QAnon tweets}
In this paragraph, we detail the keywords used to identify messages related to the QAnon conspiracy theory. The used keywords are the hashtags listed as follows:\\
 \#qanon, \#8kun, \#qpatriots, \#q, \#qarmy, \#qanon2018, \#qanon2019, \#qwarriors,  \#thegreatawakening, \#wwg1wga, \#hollywoodanon, \#qteam, \#qanons, \#qclearance, \#qbaby, \#pedogate, \#pizzagate, \#darktolight,\#wakeupamerica, \#thecollectiveq, \#thestorm.

\end{document}